\documentclass[sigconf,edbt]{acmart-edbt2021}

\def\BibTeX{{\rm B\kern-.05em{\sc i\kern-.025em b}\kern-.08em
    T\kern-.1667em\lower.7ex\hbox{E}\kern-.125emX}}

\usepackage{booktabs} %

\usepackage{mleftright}

\newif\ifextended
\extendedtrue

\usepackage{listings}%
\definecolor{codegreen}{rgb}{0,0.6,0}
    \definecolor{codegray}{rgb}{0.5,0.5,0.5}
    \definecolor{codepurple}{HTML}{C42043}
    \definecolor{backcolour}{HTML}{F2F2F2}
    \definecolor{bookColor}{cmyk}{0,0,0,0.90}  

    \lstdefinestyle{mystyle}{
        keywordstyle=\color{codepurple},
        stringstyle=\color{codepurple},
        basicstyle=\footnotesize,
        breakatwhitespace=false,         
        breaklines=true,                 
        keepspaces=true,                 
        showspaces=false,                
        showstringspaces=false,
        showtabs=false,      
    }
\usepackage{graphicx}
\usepackage{balance}
\usepackage[disable]{todonotes}
\usepackage{comment}
\usepackage{subcaption}
\usepackage{hyperref}
\usepackage{amsmath}
\usepackage{enumitem}
\usepackage{array}%
\usepackage[ruled,vlined]{algorithm2e}

\setcopyright{rightsretained}

\acmDOI{.}

\acmISBN{TBC}

\acmConference[x]{x}{.}{..} 
\acmYear{2021}

\renewcommand\footnotetextcopyrightpermission[1]{} 

\begin{document}
\title{COAX: Correlation-Aware Indexing\\ on Multidimensional Data with Soft Functional Dependencies}

\author{Ali Hadian\textsuperscript{1}, %
Behzad Ghaffari\textsuperscript{1}, %
Taiyi Wang\textsuperscript{2}, %
Thomas Heinis\textsuperscript{1}}
\orcid{0000-0003-2010-0765}
\affiliation{%
  \institution{\textsuperscript{1}Imperial College London, \textsuperscript{2}Johns Hopkins University}
}

\renewcommand{\shortauthors}{A. Hadian et. al.}

\begin{abstract}

Recent work proposed learned index structures, which learn the distribution of the underlying dataset to improve performance. The initial work on learned indexes has shown that by learning the cumulative distribution function of the data, index structures such as the B-Tree can improve their performance by one order of magnitude while having a smaller memory footprint.

In this paper, we present COAX, a learned index for multidimensional data that, instead of learning the distribution of keys, learns the correlations between attributes of the dataset. Our approach is driven by the observation that in many datasets, values of two (or multiple) attributes are correlated. COAX exploits these correlations to reduce the dimensionality of the datasets.

More precisely, we learn how to infer one (or multiple) attribute $C_d$ from the remaining attributes and hence no longer need to index attribute $C_d$. This reduces the dimensionality and hence makes the index smaller and more efficient. 

We theoretically investigate the effectiveness of the proposed technique based on the predictability of the FD attributes. We further show experimentally that by predicting correlated attributes in the data, we can improve the query execution time and reduce the memory overhead of the index. In our experiments, we reduce the execution time by 25\% while reducing the memory footprint of the index by four orders of magnitude. 

\end{abstract}

\maketitle

\section{Introduction}

Multidimensional data plays a crucial role in  data analytics applications.%
Indexing multidimensional data, however, is challenging as the curse of dimensionality hits: with every additional dimension indexed, the performance of the index degrades. %
A promising approach to tackle the curse of dimensionality is using machine learning techniques and exploit patterns in data distribution for more efficient indexing. For example, learned indexes automatically model and use the distribution of the underlying data to locate the data records~\cite{kraska2018case, galakatos2018tree,kipf2020radixspline, nathan2020learning, ding2020alex, hadian2019interpolation} and the idea has been extended to indexing multidimensional data~\cite{nathan2020learning}.

In this paper we develop a new class of learned indexes for multidimensional data that uses dependencies between attributes of either the full or a subset of the dataset to improve performance of the index structure. Our approach is motivated by the observation that it is common in real-world datasets for two or more attributes of the data to correlate. 
We argue that by taking learned indexes to the multidimensional case, in addition to learning from CDF of the data, we can also learn from relationships between attributes of the data such as the correlation between $id$ and $timestamp$, a common case in real-world datasets; or between flight distance and flight time in an airline dataset. The idea of learning the relationship between attributes of data has already been used to improve estimating the selectivity of a given query and thus to improve query optimizers~\cite{ilyas2004cords} and one-dimensional secondary indexes~\cite{wu2019designing}. We take this idea a step further to indexing multidimensional data.

More precisely, we develop models that explain correlations in the attributes of the data and we show that a multidimensional index does not need to index every dimension if some attributes are correlated. The index only needs to store one dimension per each group of correlated attributes, thereby effectively reducing the dimensionality of the dataset. In case a query targets an attribute $C_d$  that is not indexed but there is another indexed attribute $C_x$ correlated to $C_d$ ($C_x\rightarrow C_d$), we use our model to check which range of values in $C_x$ correlate with the query and run a translated query over the indexed attribute instead. As we show experimentally, the suggested approach significantly shrinks the memory footprint of the index, by four orders of magnitudes depending on the number of the FDs and their degree of correlation, while improving the overall lookup time of the indexes by 25\%.

\section{Related Work}
\label{sec:relatedwork}
Our ideas build on model-based indexes, including learned indexes and interpolation search, as well as spatial indexes. 

Recently, it was suggested that models driven from data can improve the lookup time of indexes and reduce memory footprint in what are called learned indexes~\cite{kraska2018case}. In a learned index, the CDF of the key distribution is learned by fitting a model; then the learned model is used as a replacement for the conventional index structures (such as B+trees) for finding the location of the query results. Index learning frameworks such as the RMI model~\cite{kraska2018case} are capable of learning arbitrary models, although a further theoretical study~\cite{ferragina2020learned} %
have shown that models such as linear splines are effective for most real-world datasets. Spline-based learned indexes include Piecewise Geometric Model index (PGM-index)~\cite{ferragina2020pgm}, Fiting-tree~\cite{galakatos2018tree}, Model-Assisted B-tree (MAB-tree)~\cite{hadian2020madex}, Radix-Spline~\cite{kipf2020radixspline}, Interpolation-friendly B-tree (IF-Btree)~\cite{hadian2019interpolation} and others
~\cite{llaveshi2019accelerating,setiawan2020function}. A comprehensive comparison of learned indexing approaches can be found here~\cite{ferragina2020survey}.

COAX is inspired by hybrid learned indexes that combine machine learning with traditional indexes structures, including RadixSpline~\cite{kipf2020radixspline}, FITing-tree~\cite{galakatos2018tree}, Interpolation-Friendly B-tree~\cite{hadian2019interpolation}, and MADEX~\cite{hadian2020madex}. Hybrid learned indexes have been well explored in the one-dimensional case. 
\ifextended
For example, RadixSpline~\cite{kipf2020radixspline} uses radix-trees as the top-level model while FITing-tree~\cite{galakatos2018tree} and IFB-tree~\cite{hadian2019interpolation} use B+-tree as the top-level index structure. Then they use a series of piecewise linear functions in the leaf level. 
\fi
Furthermore, adaptability and updatability of learned indexes have been explored in a similar area \cite{ding2020alex, hadian2019considerations}. Finally, in the multivariate area, learning from a workload has also shown interesting results \cite{nathan2020learning,hadian2020handsoff}.

\ifextended
Besides the main trend in learned indexes (focusing on range indexing), machine learning has also inspired other indexing and retrieval tasks. This includes bloom filters~\cite{mitzenmacher2018model,dai2019adaptive}, inverted indexes~\cite{xiang2018pavo}, conjunctive Boolean intersection\cite{oosterhuis2018potential}, learned indexing for strings~\cite{wang2020sindex}, rule-matching~\cite{li2019accelerating,zhang2020efficient}, and computing list intersections~\cite{ao2011efficient}.
\fi

Another direction that we exploit in this paper is the statistical dependency between the attributes (data columns), i.e.,  the notion of \textit{soft functional dependencies (Soft FDs)}~\cite{ilyas2004cords,wu2019designing}, which is a relaxed definition of FDs. A soft FD between X and Y means that the value of X determines the value of Y with a high probability, i.e., we can predict Y using X but there might be errors in prediction, for example, from Flight departure time we can determine the approximate arrival time.
Soft FDs have been mainly used for selectivity estimation in query optimization~\cite{ilyas2004cords,kimura2009correlation,liu2016exploiting}, designing materialized views and indexes~\cite{zdonik2010coradd}, data integration from multiple data sources~\cite{wang2009functional}, and reducing the number of secondary indexes~\cite{wu2019designing}. In this work, however, we aim to design an index structure that uses soft FDs to decrease memory footprint and accelerate query execution in a \textit{multidimensional index}.

In contrast to the related work like HERMIT, which is based on unclustered secondary indexes, COAX builds a multidimensional primary index, which is more efficient for range indexing as it does not require heavy join operations for multidimensional queries. Moreover, related work only considers cases with a small percentage of outliers (the records that do not follow the dependency pattern) while COAX consists of two indexes where the outliers are organised as a separate multidimensional index, hence supporting datasets with a much `softer' functional dependency, i.e., a considerable number of outliers (e.g., $25\%$). Finally, previous work on soft-FD requires the DBA to manually introduce all soft-functional dependencies and handcraft the index layout for the soft-FD, which is very time-consuming for the DBA. COAX, on the other hand, automatically detects the correlated columns and works with any multidimensional index structure.

\section{Approach Overview}

\label{sec:overview}
We are interested in exploiting correlations in a multidimensional dataset to build a more efficient index.
The key components of COAX, illustrated in Figure~\ref{fig:overview}, are as follows:

\begin{figure}
\centering
    \includegraphics[width=.99\linewidth]{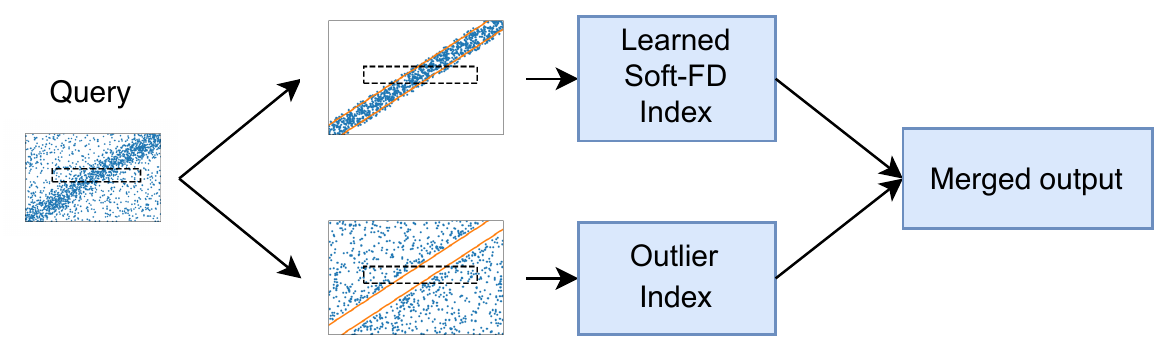}
\vspace{-7pt}
\caption{The suggested index structure: a primary index (top) indexes only one dimension from each group of correlated attributes, while a small outlier index handles the outliers and guarantees retrieval of all results.} 
\label{fig:overview}
\vspace{-10pt}
\end{figure}

\begin{itemize}[leftmargin=*]
    \item \textbf{Query translation.} 
    The core idea of COAX is that to build an $n$-dimensional index on a data where $m$ attributes are highly correlated with the other $n-m$ attributes (i.e., $m$ dependent attributes), we can build a multidimensional index on the $n-m$ attributes which results in a much smaller multidimensional index.  If a query constraint targets one of the dependent attributes, COAX transforms the constraint into another constraint on one of the indexed attributes. Query translation uses a model that predicts the \textit{dependent attributes} using a set of \textit{indexed attributes}, along with the tolerance margins. 
    \item \textbf{Learning the correlations.} COAX detects whether a functional dependency exists between two or more attributes, and evaluates whether the dependency can be effectively modelled in the presence of outliers.
    \item \textbf{Learned Soft-FD index.} We apply a pre-processing step to our data. We use the learned correlations to separate data that agrees with the learned dependency from the remaining points that are regarded as outliers. We create a \textit{primary index} on the data points that have up to a certain deviation from the learned correlation (i.e., those that are within a certain tolerance margin around the fitted line). The primary index only indexes one attribute per each set of correlated attributes. For the rest of the correlated attributes, a model is learned to represent the correlation between \textit{indexed attributes} and \textit{learned attributes} is used to execute queries. If a query targets a dependent attribute $C_d$ that is not indexed but is correlated with an indexed attribute $C_x$, then the learned correlation model converts query constraints targeting $C_d$ to an equivalent constraint on $C_x$. 
    \item \textbf{Outlier index.} Data points that do not fall within the tolerance margin of the soft FD model are excluded from the primary index and are indexed with all dimensions in an \textit{outlier index}, which is a typical multidimensional index structure.
\end{itemize}

\section{COAX Query Translation}
\label{sec:query-translation}
Let us consider the simple case where we want to answer queries on two attributes $C_1$ and $C_2$:

\begin{lstlisting}[language=SQL,
morekeywords={clustered},    
mathescape=true]
    SELECT * FROM tbl WHERE     $q_1^{low} < C_1 < q_1^{high}$ 
                        AND     $q_2^{low} < C_2 < q_2^{high}$; 
\end{lstlisting}

We define a query by a rectangle characterised by its lowermost leftmost point $(q_1^{low},~ q_2^{low})$ and uppermost rightmost point $(q_1^{high},~ q_2^{high})$. Note that with this setting, we can express the case where, for example, only the first dimension is queried by defining $q_2^{high}=\infty$ and $q_2^{low}=-\infty$. Similarly, we can express point queries by defining the lower and upper points to be equal. %

Suppose that the attributes to be indexed are correlated, i.e., there is a soft functional dependency between one attribute, say $C_x$ (the \textit{indexed} attribute), and another dependent attribute $C_d$. 
In this case, a range index can be built on only one of the attributes ($C_x)$, and the query constraint that targets a non-indexed attribute $C_d$ can be mapped to equivalent query constraints on the indexed attribute using the prediction model and its error bounds. To do this without loss of accuracy, we need an oracle dependency function $\psi: C_x \rightarrow C_d$ that calculates the value of $C_d$ based on a given $C_x$ for each row in the dataset. Because in practice finding such an oracle function may prove impossible for a \textit{soft} functional dependency, we must relax this concept and allow our model to instead estimate an approximate value for $C_d$, i.e., $\hat{\psi}: C_x \rightarrow C_d$

Similar to learned range indexes \cite{kraska2018case}, using an approximation without any bounds is impractical since we want to avoid scanning the entire dataset. We must therefore define a tight error bound for the approximation. 
To do so, we argue that once a significant majority of the values of the dependent attribute $C_d$ are very close to the values predicted by $\hat{\psi}$ model. For any point $(p_x, p_d) \in (C_x, C_d)$ in our primary index we have:

\begin{equation}
p_d \in [\hat{\psi}(p_x) - \epsilon_{LB}, \psi(p_x) + \epsilon_{UB}]
\end{equation}

Where $\epsilon_{LB}$ and $\epsilon_{UB}$ are the the lower bound and upper bound error margins, or as illustrated graphically, the distances at which the data separators have been drawn in both directions. We then keep the records that fall between these bounds and leave outliers aside to be inserted into the outlier index. Figure~\ref{fig:overview} shows the process of defining the two parallel lines characterising the lower and upper error bounds of a linear regression model, and the role of our primary and outlier index. 

\begin{figure}
  \centering
  \begin{minipage}[b]{0.44\linewidth}
    \includegraphics[width=\linewidth]{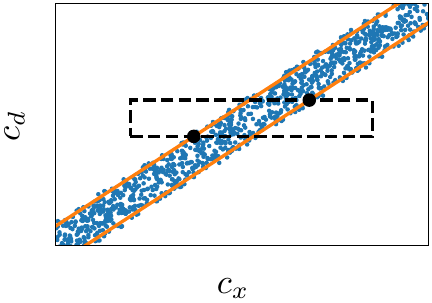}
  \end{minipage}
  \hspace{0.02\linewidth}
  \begin{minipage}[b]{0.44\linewidth}
    \includegraphics[width=\linewidth]{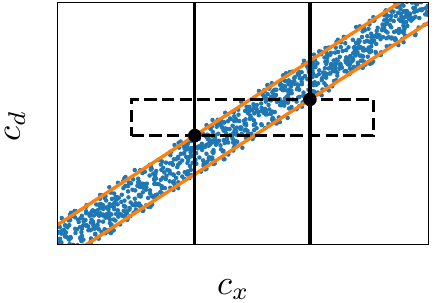}
  \end{minipage}
  \vspace{-10pt}
  \caption{Query execution on the primary index. Query constraints targeting the dependent attributes $C_d$ are mapped to equivalent constraints on the indexed attribute $C_x$ using the model and the error bounds. The final query constraint is the intersection of the two constraints on $C_x$}
  \vspace{-8pt}
\label{intersection}
\end{figure}

To create a primary index on $C_x$ and $C_d$, we only need to sort the rows based on the $C_d$ attribute. To answer a range query on both dimensions, we need to (1) calculate which range of values in the indexed dimension corresponds to the queried range for the missing dimension and (2) scan the intersection of the queried ranges projected on the indexed dimension.
Put differently, because the data in the primary index fits in within the lower and upper bound estimates of the model, we can tighten the lower and upper bounds of our query for each of the correlated dimensions making the overall scanned range smaller. As illustrated in Figure~\ref{intersection}, we need to find the intersection of the query rectangle with our lower and upper thresholds. We then scan the records in the indexed attribute ($C_d$) between the more selective bounds (drawn as solid vertical lines in Figure~\ref{intersection}), i.e.:
\begin{equation}
\hspace{-5pt}
\left[max\left(\hat{\psi}(q_1^{low}), q_2^{low}-\epsilon_{LB}\right), min\left(\hat{\psi}(q_1^{high}), q_2^{high}+\epsilon_{UB}\right)\right]
\end{equation}

\section{Training the Soft-FD Models}
\label{sec:detecting-the-dependencies}

Detecting the soft-FDs and learning a soft-FD model that 
is computationally expensive. To tackle this, COAX uses a fraction of keys rather than the full key set to train the model and evaluate the efficiency of adopting a potential soft-FD.
More precisely, COAX only considers centres of \textit{dense areas} in a sample drawn from the dataset. We overlay a multidimensional grid on the key space and count the number of records in each cell. We then filter out any cells that do not reach a threshold in their count and consider our training data to be the weighted centres of the remaining cells.
\ifextended
This is depicted in Figure~\ref{fig:densepoints}. %

\begin{figure}
  \centering
  \begin{minipage}[b]{0.44\linewidth}
    \includegraphics[width=\linewidth]{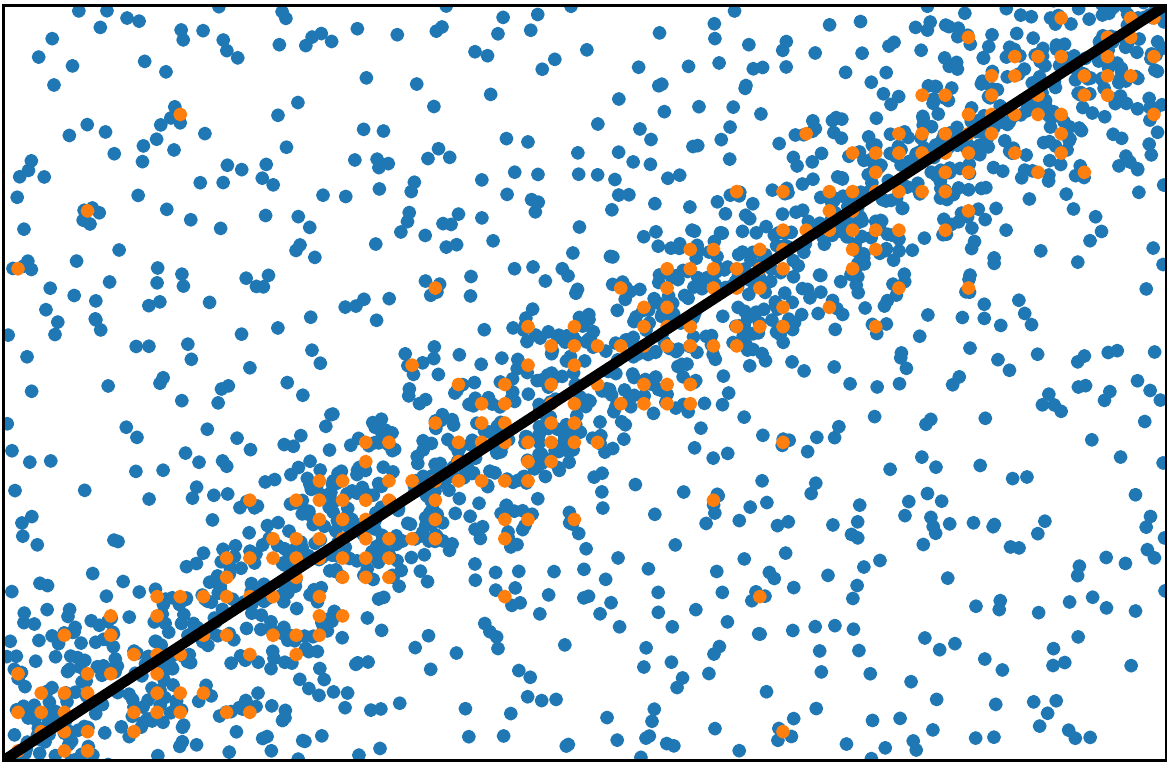}
  \end{minipage}
  \hspace{0.03\linewidth}
  \begin{minipage}[b]{0.44\linewidth}
    \includegraphics[width=\linewidth]{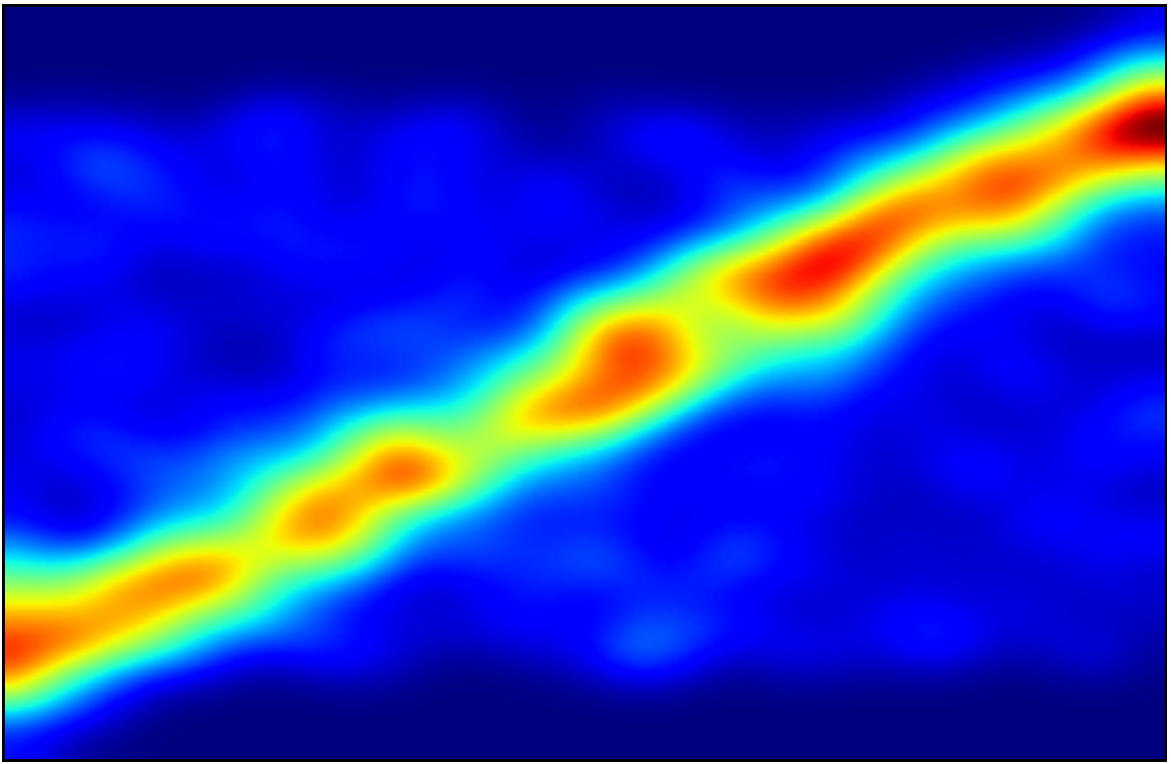}
  \end{minipage}
  \caption{An error margin can be defined by considering the  the correlation model of the data (left) and the density of the data records around the model (right)}
  \label{fig:densepoints}
\end{figure}
\fi

After the training set (bucket centres) is found, we recursively consider unique pairs of attributes and use a Monte Carlo sampler to check whether a linear model fits the training records (Algorithm \ref{alg:split}) to learn correlations between multiple attributes. If two attributes are found to be correlated, we save the resulting pair along with their model parameters. In the final step, we merge all groups that have an attribute in common and pick one attribute in each group to be the predictor responsible for estimating the remaining attributes in its group. %

\begin{algorithm}
\SetAlgoLined
\textbf{Input:} Centred attribute values: $C_x$, $C_d$ [$N$] \\
\textbf{Result:} Model parameters ($m$, $b$), $primary\_index$, $outlier\_index$ \\
$C_x\_sample$ = $C_x$.sample($sample\_count$) \\
$C_d\_sample$ = $C_d$.sample($sample\_count$) \\
$w_x$ = $C_x\_sample$.max() / $bucket\_chunks$ \\
$w_d$ = $C_d\_sample$.max() / $bucket\_chunks$ \\
$buckets$ = [$bucket\_chunks$][$bucket\_chunks$] \\
\For{$i\gets 0$ \KwTo $sample\_count$}{
    $buckets$[$C_x$[$i$] / $w_x$][$C_d$[$i$] / $w_d$] += $1$ \\
}
$C_x\_train$ = $C_d\_train$ = [] \\
\For{$i\gets 0$ \KwTo $bucket\_chunks$}{
    \For{$j\gets 0$ \KwTo $bucket\_chunks$}{
        \If{$buckets$[$i$][$j$] $>$ $threshold$} {
            $C_x\_train$ += [i * $w_x$ + 0.5$w_x$] * $buckets$[$i$][$j$] \\
            $C_d\_train$ += [j * $w_d$ + 0.5$w_d$] * $buckets$[$i$][$j$] \\
        }
    }
}
$m, b$ = linear\_regress($C_x\_train$, $C_d\_train$) \\
$displacements$ = $(C_d - (m \cdot C_x - b))$\\
\For{$i\gets 0$ \KwTo $N$}{
    \eIf{$- \epsilon_{LB} < displacements[i] < \epsilon_{UB}$} {
        $primary\_index$.insert($C_x$[$i$], $C_d$[$i$]) \\
    } {
        $outlier\_index$.insert($C_x$[$i$], $C_d$[$i$]) \\
    }
}

\caption{Splitting Data} 
\label{alg:split}
\end{algorithm}

The accuracy and runtime of the learning step can be adjusted by tuning parameters given in our proposed algorithm. To increase the accuracy of the model, we can draw a larger learning sample from the dataset, group them in smaller cells and define a lower cell acceptance threshold. This, however, increases the number of records processed by the regression algorithm and can degrade training run time. 

The bucketing step significantly reduces the number of records the regression algorithm has to consider without impeding the method's ability to generalise. This reduces the training time but more importantly, it provides an added benefit; recall that we have already inserted our sample records in a grid index (to use centre points of its denser cells to define our training set), we can maintain the trained index for later use. As new records are added, a sample of new items can be inserted in the existing grid index without having to populate it from scratch. This in combination with the fact that we have used a Bayesian method for learning the regression model, can help supporting updates on the index, as we can use the previous gradient and intersect and continuously adjust our existing model.

\section{Index Implementation}
\label{sec:indexlayout}

To evaluate differences in performance when predicting dimensions, we need to build an index on \textit{indexed dimensions}. We implement our index on top of Grid Files~\cite{nievergelt1981grid} with a number of modifications. In particular, we choose boundaries for each cell based on quantiles along each dimension and use the same number of grid lines for each attribute. Addresses for all cells are sorted using the original ordering of attributes in the dataset. Furthermore, each cell stores records in a contiguous block of virtual memory in a row store format. Finally, rows within each page are sorted based on a given function similar to the approach proposed in Flood \cite{nathan2020learning}. Sorting the rows inside pages means that we can reduce the dimensionality of the grid by one. This is because instead of having grid lines for the particular sorted attribute, we can use binary search to locate items (or a scan between two bounding binary searches in a range query).

\ifextended

Note that picking grid lines based on the distribution of the dataset does not mean that we have regular cells. Although doing so does reduce the standard deviation in cell lengths for non-uniform datasets, bucket lengths are still allowed to grow arbitrarily large. Figure~\ref{fig:histofpagesize} shows the variation in cell lengths in one of our experiments.

\begin{figure}
  \centering
    \begin{minipage}[b]{\linewidth}
  \centering
    \includegraphics[width=.75\linewidth]{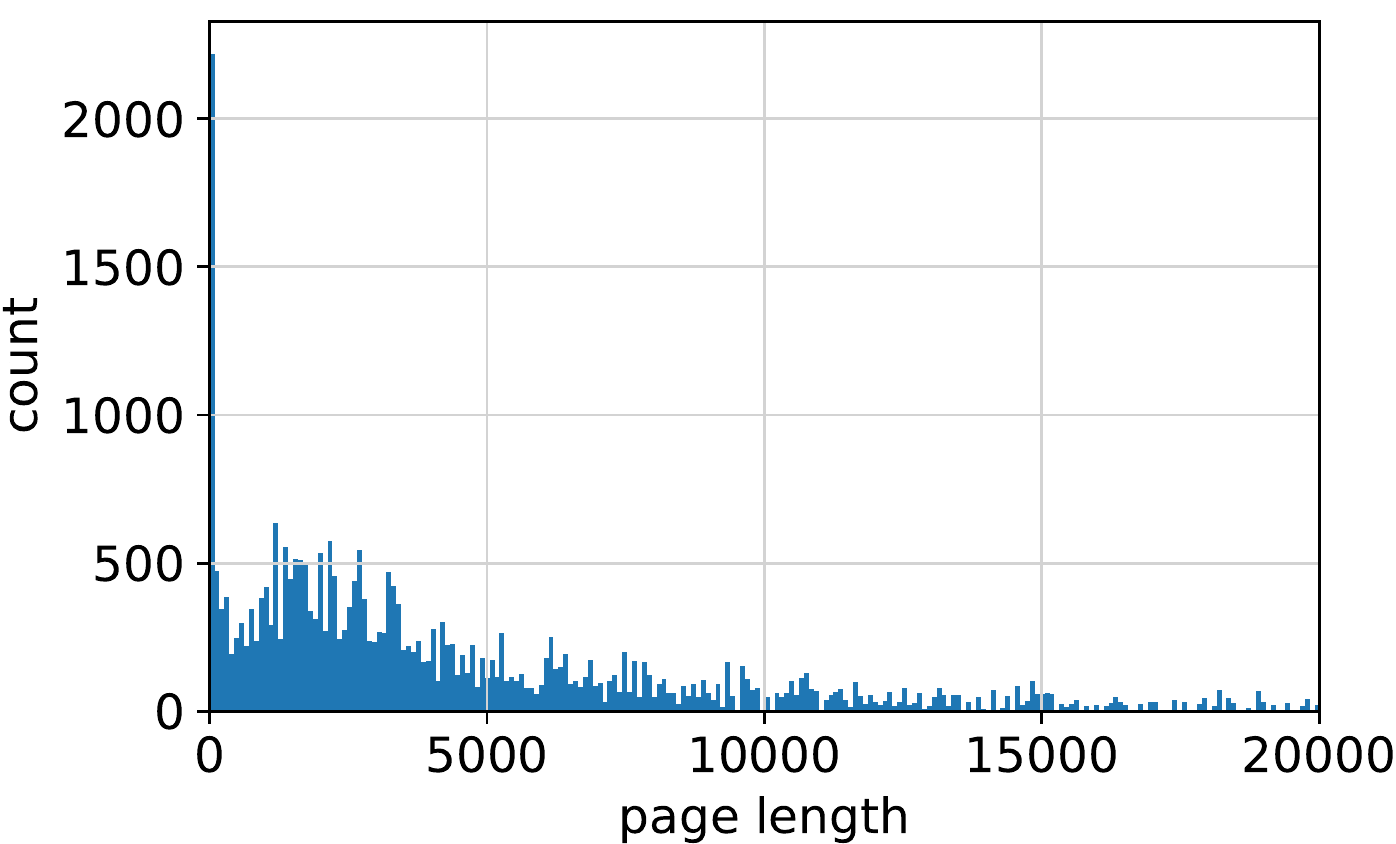}
    \subcaption{Non-uniform distribution of page sizes in 2D grid layout}
    \label{fig:histofpagesize}
  \end{minipage}
  \hfill
  \begin{minipage}[b]{0.46\linewidth}
    \includegraphics[width=\linewidth]{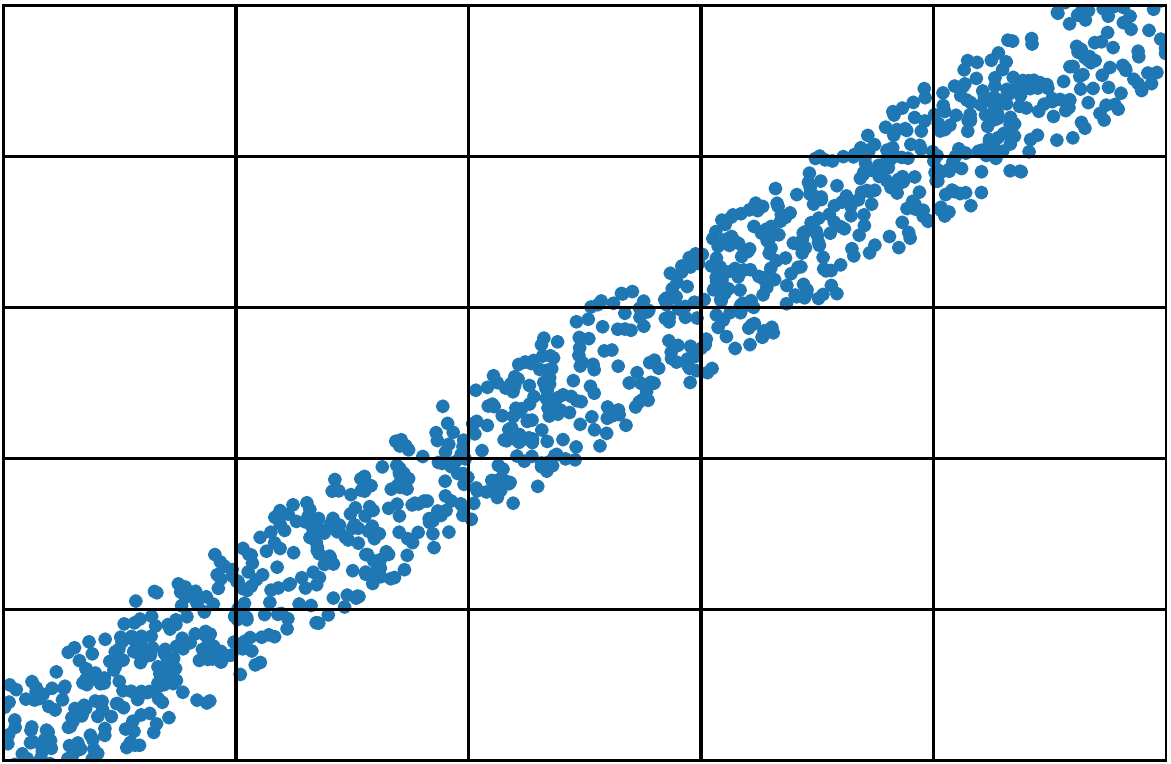}
    \subcaption{2D Index Layout}
  \end{minipage}
  \hfill
  \begin{minipage}[b]{0.46\linewidth}
    \includegraphics[width=\linewidth]{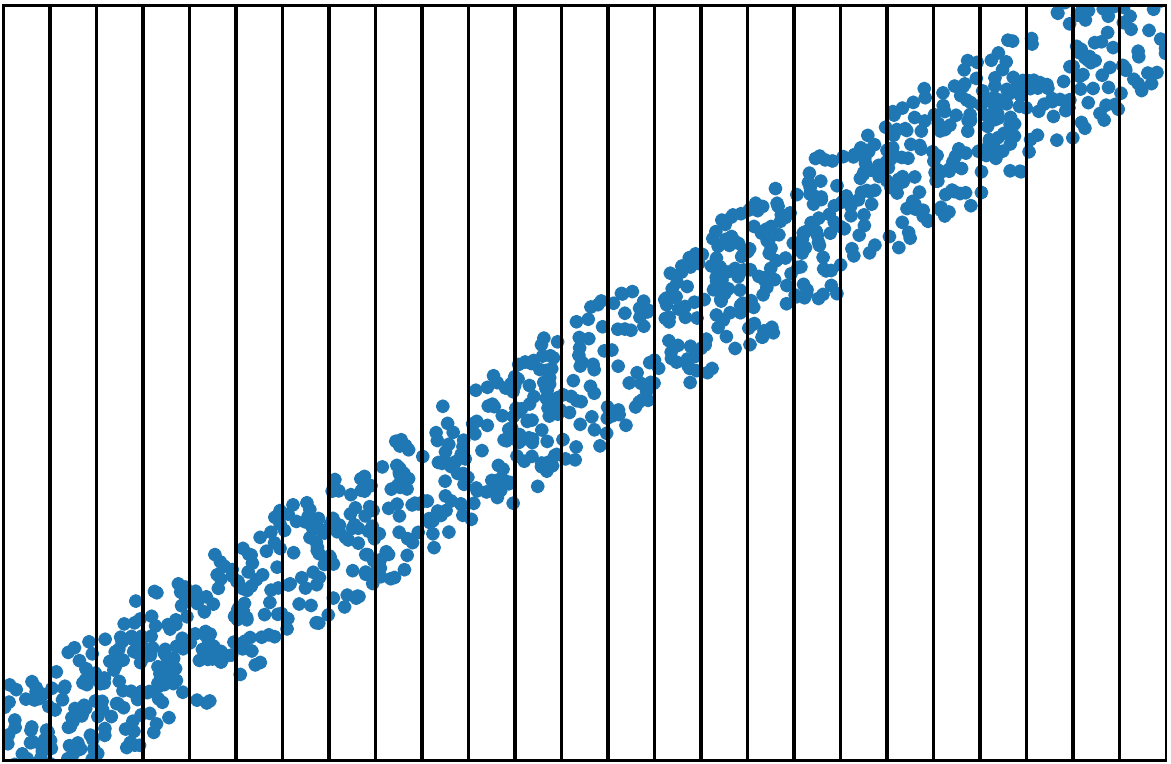}
    \subcaption{Learned 1D Grid}
  \end{minipage}
  \hfill
  \caption{Reducing index dimensionality allows having more accurate grids for the remaining predictor dimensions}
  \vspace{-10pt}
 \label{gridshape}
\end{figure}

\fi

The combination of predicting attributes and having a sorted dimension means that for a dataset with $n$ dimensions and $m$ predicted attributes, we only need an index with $n-m-1$ dimensions. Due to this reduced dimensionality, we will show in section \ref{sec:memoryperformancetradeoff} that the resulting index makes much more efficient use of memory.

\section{Theoretical Analysis}
\label{sec:theoretical_analysis}
In this section, we analyze how the performance of the soft-FD model is affected by the margin size and how its memory footprint compares with a multidimensional index (square grid) that performs the same number of comparisons.

\subsection{Effect of Margin Size}
\begin{figure}
\centering
\includegraphics[width=0.99\linewidth]{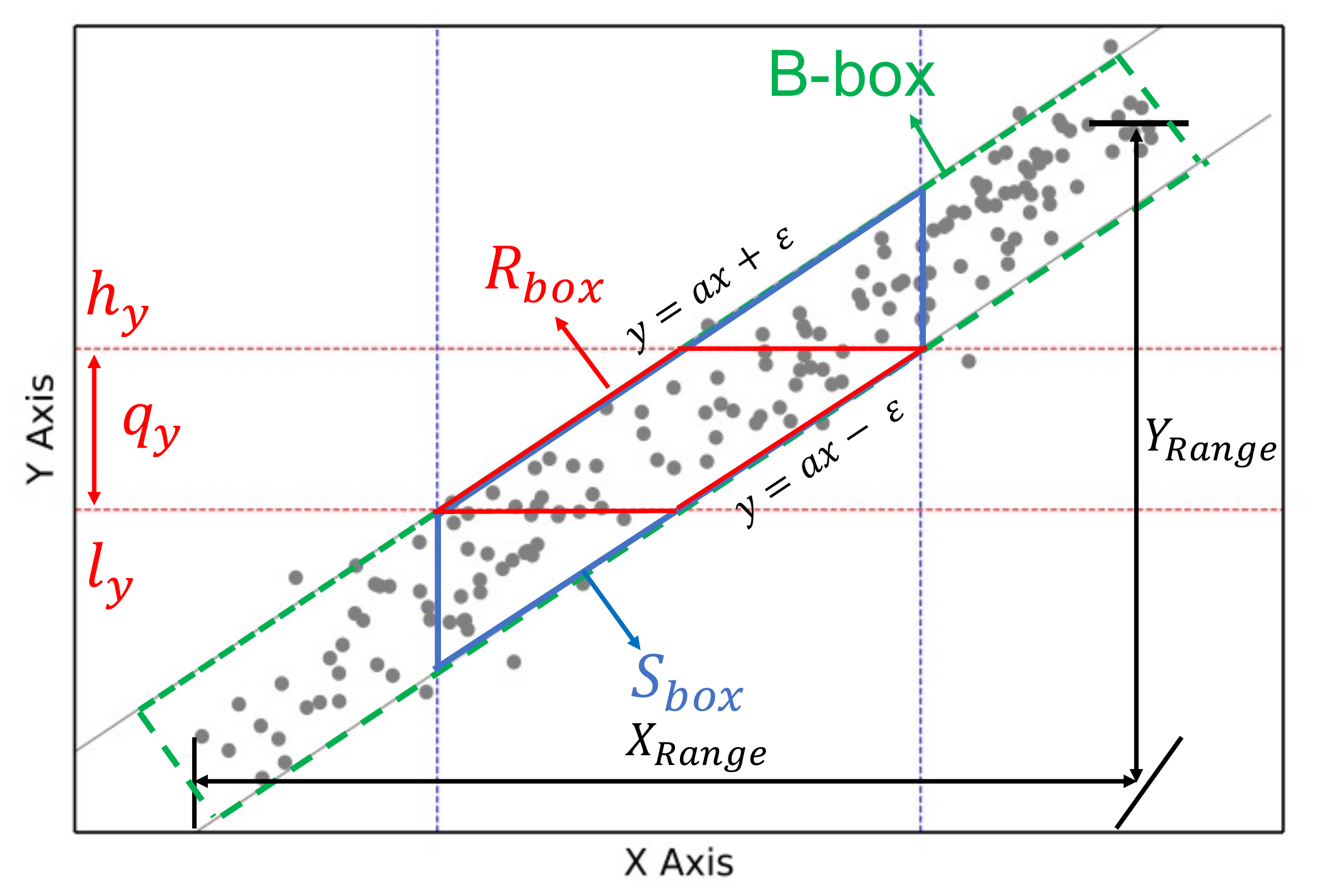}
\vspace{-10pt}
\caption{Illustrating the notation. B-box defines the boundaries of the index (the range of records indexed by the Soft-FD index). R-box (red parallelogram) is the result set. S-box (blue parallelogram) is the area scanned. The range query $(h_y,l_y)$ is shown with red dashed lines.}
\label{fig:theory_softfd_scan_areas}
\end{figure}

Figure~\ref{fig:theory_softfd_scan_areas} compares the area scanned by the soft-FD model, with the area of the actual result set. We define the following regions:

\begin{itemize}[leftmargin=*]
    \item \textit{B-box} (shown as a green box) is the index \textit{ boundaries}): the area that contains all data points (records) restricted by two borderlines and hence supported by the Soft-FD index. Any data point outside of B-box needs to be handled by the outlier index.
    \item \textit{R-box} is the \textit{Result set}, which is shown as a red parallelogram.
    \item S-Box is the \textit{Scanned} area, illustrated by a  blue parallelogram. For the index to return all results, we need have $\textit{R-box} \subset \textit{S-box}$. Both S-box and R-box are within the green rectangle with dashed lines.
    
\end{itemize}

Without loss of generality, we assume that the margin has the same length from both sides, i.e., $\epsilon_{LB} = \epsilon_{UB} = \epsilon$. Also, we consider linear models only, and assume that B-box is symmetric, which means that $\hat{\psi}(x) = ax$. Therefore, the two boundary lines are defined as $ y=ax+ \varepsilon$ and $y= ax- \varepsilon$. Note that, if the model is non-symmetric ($\hat{\psi}(x) = ax + b, b\neq 0$) or if $\epsilon_{LB} = \epsilon_{UB}$, the areas of B-box, S-Box, and R-box are the same and we transform the problem to a symmetric equivalent with a parallel movement.

The query constraints can be on the X-axis, the Y-axis, or both. The constraints on either of the axes are translated to its equivalent in the other axis (considering the $\pm \varepsilon$ boundaries) and eventually define some S-box parallelogram as explained in Section~\ref{sec:query-translation}. Therefore, without loss of generality, here we consider a range query with constraints on the Y-axis only while other queries can be translated to this case. Given the pre-defined range $(h_y,l_y)$, i.e., we want to find all points $(x_i,y_i)$ where $y_i \in [q_y,l_y]$. Let us denote the data range on the X-axis (the indexed attribute) as $X_{range}$ and the range on the Y-axis (the dependent attribute) as $Y_{range}$.

We define $q_y$ as the range of the query $q_y = h_y - l_y$. The area of R-box (the red parallelogram) is denoted as $S_r$. R-box is the result set, and we ideally expect the soft-FD index to scan exactly the R-box region (i.e., the red parallelogram in Figure~\ref{fig:theory_softfd_scan_areas}). 
 Following the equations of boundary lines  $ y=ax+ \varepsilon$ and $y= ax- \varepsilon$ and the definition of $\varepsilon$ above, $S_r$ can be computed using the base of the R-Box ($Base_{r}$) and its height ($Height_{r}$):

\begin{equation}
\label{eq:s_r}
S_r = Base_{r} \cdot Height_{r} =(\frac{hy+\varepsilon}{a}-\frac{hy-\varepsilon}{a}) \cdot q_y = q_y \cdot \frac{2\epsilon}{a}
\end{equation}

For the Soft-FD model, we define the scanned area $S_s$ as the area of the S-box (the blue parallelogram in Figure~\ref{fig:theory_softfd_scan_areas}). $S_s$ is likewise estimated using the base and height of the S-box:

\begin{equation}
    S_s = Base_{s} \cdot Height_{s} = \frac{2\varepsilon(2\varepsilon+q_y)}{a}
\end{equation}

We define the \textit{effectiveness} of the Soft-FD model as the ratio between $S_r$, the effective scanning area in ideal cases, and $S_s$, the actual area that the Soft-FD index needs to scan:

\begin{equation}
\label{eq:effectiveness}
   \text{effectiveness} = \frac{S_r}{S_s} = \frac{q_y}{2\varepsilon+q_y}
\end{equation}

Effectiveness is the ratio of the red area and blue area in Figure~\ref{fig:theory_softfd_scan_areas}, which is affected by the width of the B-box ($2\epsilon$). It can be concluded from Equation~\ref{eq:effectiveness} that when the margin is tighter and $\varepsilon$ gets closer to 0, effectiveness gets closer to 1, which means the area that the Soft FD index estimates for scanning match the actual result set, i.e., S-Box fits R-Box and $S_r = S_s$. Similarly, when $\varepsilon$ gets larger, the ratio gets smaller, which means that the Soft-FD model is less effective. We assumed that the data distribution is close to uniform, otherwise the PDF of the distribution has to be considered in Equation~\ref{eq:effectiveness}.

\subsection{The Capacity of Linear Soft-FD Models}
\label{subsec:theory-capacity}
In this section, we provide a stochastic analysis to estimate the capacity of a soft-FD index. If the segmentation number n is large enough, the number of data points in the original data set and the number of keys in the new sequence are of the same order of magnitude. Thus, we can estimate the capacity of a soft-FD index with the properties of the new sequence. Our key results are:
\begin{itemize}[leftmargin=*]
    \item The expected number of keys covered by a linear segment (Theorem \ref{t1})
    \item The number of segments needed to cover a stream of length n (Theorem \ref{t4}), which is the capacity of a linear model for a soft-FD index. 
\end{itemize}

We also provide additional conclusions, including  Theorem \ref{t2} which helps us explain the necessary conditions for the optimal value of Theorem \ref{t1}. %
\ifextended
The proofs of all theorems is provided in the Appendix.
\else
The proofs of all theorems is provided in the Appendix of the extended version of this paper\footnote{\url{https://arxiv.org/abs/2006.16393}}.
\fi

Ferragina et al.~\cite{ferragina2020learned}  suggest a stochastic process to measure the capacity of learned indexes by making an approximate assumption of a stochastic process for the position-key sequence and designed reasonable distributions over X-axis (Key axis). We build on their thinking with the help of our CSM model, and we can directly assume the gap distribution of the Y-axis data, so that we no longer need the swapping of the coordinate. In other words, by discretizing the continuous value-value sequence via the CSM model, we generalized their stochastic sequence model.  In practice, we assume that the sequence gaps $\{g_i\}$ are random variables which are sampled from the i.i.d. distribution $G_i$ with probability density function $f_G$, and then we can make assumptions on the mean and variance of the gap distribution, with mean $\mathbb{E}[G_i] = \mu$ and variance $Var[G_i] = \sigma^2$.

\begin{theorem}
In the soft-FD model, if a single linear segment has slope $a=\mu$ and the margin parameter $\varepsilon$ is sufficiently larger than the variance of the gap distribution $\sigma$, the expected number of keys covered by this linear segment is
$
 \frac{\varepsilon^2}{\sigma^2}
$
\label{t1}
\end{theorem}

\begin{theorem}
Given the same assumptions of Theorem \ref{t1}, the expected number of keys covered by a single linear segment can be expressed as a function of margin parameter $\varepsilon$ and the variance of the gap distribution $\sigma$. The maximum value of this function, i.e. the largest expected number of keys covered is achieved when the slope is $\mu$.
\label{t2}
\end{theorem}

\begin{theorem}
Given the assumptions and symbols of Theorem \ref{t1}, the variance of the number of keys covered by a single linear segment is $\frac{2\varepsilon^4}{3 \sigma^4}$ 
\label{t3}
\end{theorem}

Even though this paper mainly discusses a linear soft-FD model (in terms of a linear model with a margin of constant width), one can use more complicated non-linear methods to model soft functional dependencies that cannot be modelled with a linear model. Among non-linear models, we specifically consider linear splines, which are recently shown to be very effective in learned indexes for modelling the value-to-position mappings~\cite{galakatos2018tree,kipf2020radixspline,ferragina2020pgm}. Linear splines still consist of linear models, where each linear model is used in a specific region. Theorem~\ref{t4} suggests the number of segments that a linear spline model requires for modelling soft-FDs with a maximum error of $\epsilon$. 

\begin{theorem}
Following the basic assumptions of Theorem \ref{t1}, given a data stream of length n and the margin parameter $\varepsilon$, the number of segments s needed to cover the stream can be computed from the results of Theorem \ref{t1} and Theorem \ref{t2}, which converges to
$n \cdot \frac{\sigma^2}{\varepsilon^2}$
\label{t4}
\end{theorem}

\section{Evaluation}
\label{sec:evaluation}

\subsection{Experimental Setup}
\subsubsection{Implementation and Runtime Environment} We implement the online section of our index in C and compile it using Clang 10.0.0. All of the indexes in our experiments run in a single thread and use single-precision floating point values. We build the soft-FD model using Python and the pymc3 library \cite{pymc3}. We run our experiments on an Intel Core i5-8210Y CPU running at 3.6 GHz (L1: 128KB, L2: 512KB, L3: 4MB) and 8GB of RAM.

\subsubsection{Datasets} 
We run experiments on two real-world datasets:

\textit{Open Street Map (\textit{OSM})}: we use 4 dimensions of the OSM data for the US Northeast region \cite{osm} which contains 105M records; The $Id$ and $Timestamp$ attributes in the OSM dataset are correlated and its $Latitude$ and $Longitude$ coordinates contain multiple dense areas. For this dataset we group ($Id$, $Timestamp$) for the case of learned index.

\textit{\textit{Airlines}}: data from US Airlines flights from 2000 up to 2009, which has 8 attributes and 80M records. The airline dataset is more interesting for our experiments because it contains many correlated dimensions. Example grouping in this dataset in our experiments usually consists of ($Distance$, $TimeElapsed$, $AirTime$) and ($ArrTime$, $DepTime$, $ScheduledArrTime$). We can reduce the grid dimensionality to 2-4(depending on the tolerated error threshold). Table~\ref{tbl:datasets} summarises the key aspects of the datasets.

\begin{table}
\setlength\extrarowheight{3pt}
  \begin{center}
    \label{tab:table1}
    \begin{tabular}{l c r}
     \hline
       & \textbf{Airline} & \textbf{OSM}\\
     \hline
      \textbf{Count} & 80M & 105M\\
      \textbf{Key Type} & float & float\\
      \textbf{Dimensions} & 8  &  4\\
      \textbf{Correlated Dimensions} & (3, 3)  &  2\\
      \textbf{Indexed Dimensions (Soft-FD Index)} & 2-4  &  3\\
      \textbf{Primary Index Ratio} & 92\%  &  73\% \\
     \hline
    \end{tabular}
    \caption{Dataset characteristics}
    \label{tbl:datasets}
  \end{center}
  \vspace{-15pt}
\end{table}

We generate the queries by picking a random record from the data. Then, we find the K nearest records and take the minimum and maximum values corresponding to each dimension. Our range queries are rectangles and target all attributes in the index. %

\subsubsection{Baselines} 
We compare our suggested method with the \textit{R-Tree}, arguably the most broadly used index for multidimensional data, and two grid structures as baselines: the uniform grid and column files.

\textit{Uniform grid}: or equivalently the full grid, is a hash structure that breaks down each attribute into uniformly sized grid cells between their minimum and maximum values. The address for each cell is stored independently and no adjacent cells are "shared/merged" explicitly. In memory, addresses for all cells are sorted using the original ordering of attributes in the dataset. Furthermore, each cell stores points in a contiguous block of virtual memory in a row store format.

\textit{Column files}: Essentially a non uniform grid, uses the CDF of the data to align/arrange its cell boundaries and sorts data within each cell based on one of the attributes in the data, thus reducing the dimensionality of the index by one. In a lookup on the sorted dimension, we use binary search in each cell to get the range that needs to be scanned. Column files is  similar to the approach~\cite{nathan2020learning} with the difference that it does not assume that the query workload is known and hence uses the data distribution to arrange and align the grid layout.

\textit{Full scan}: Every item in the dataset is checked against queries.

\subsection{Results}
\subsubsection{Tuning}
In this experiment we measure and compare the execution time for all indexes. We use the configuration that performs best for each index. This configuration consists of chunk size for the full grid, chunk size and sort dimension for the column files and COAX, and the node capacity (non-leaf and leaf capacity) of the R-Tree. For example, we evaluated different node capacities between 2 to 32 for the for the R-Tree index and the best performance for each experiment (i.e., point queries and individually for each selectivity rate in range queries) was used. The best node size for R-Tree is between 8 and 12.%

Due to memory constraints, we %
limit any index that would require more memory overhead for its index directory than memory occupied by the underlying data itself. %

We evaluate our results using range queries and point queries that are drawn randomly from each dataset. We define a point query as a range query where the lower bound and upper bound in the matching hyper rectangle are equal.

\subsubsection{Point and range queries}
\begin{figure}
\centering
    \includegraphics[width=\linewidth]{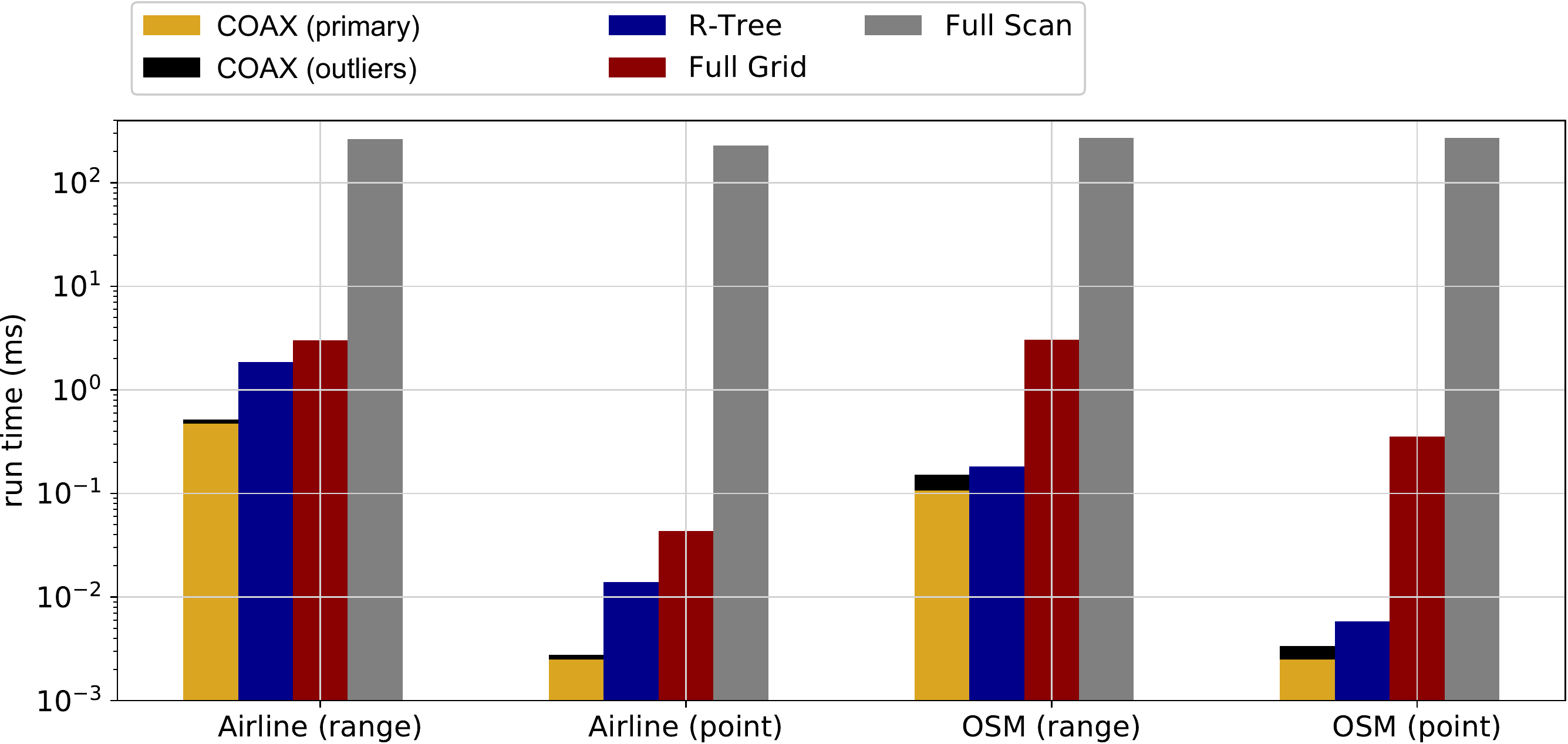}
 \caption{Query runtime performance on airline and OSM data for range and point queries. Note the log scale.}
\label{runtime}
\end{figure}

As seen in Figure~\ref{runtime}, COAX outperforms both the R-Tree and full grid. The main drawback in the case of full grid is the higher index dimensionality and the fact that it is limited in terms of how many cells it can use. 
\ifextended
An example illustration for this in 2D is shown in Figure~\ref{gridshape}. 
\fi
This is because with a skewed dataset, most grid cells become empty or very small in size. In addition, in comparison to Column Files, COAX benefits from a smaller number of memory lookups. The decreased total number of cells in COAX also translates to binary search on larger ranges in each cell, which makes COAX even more efficient.

\begin{figure}
\centering
    \includegraphics[width=\linewidth]{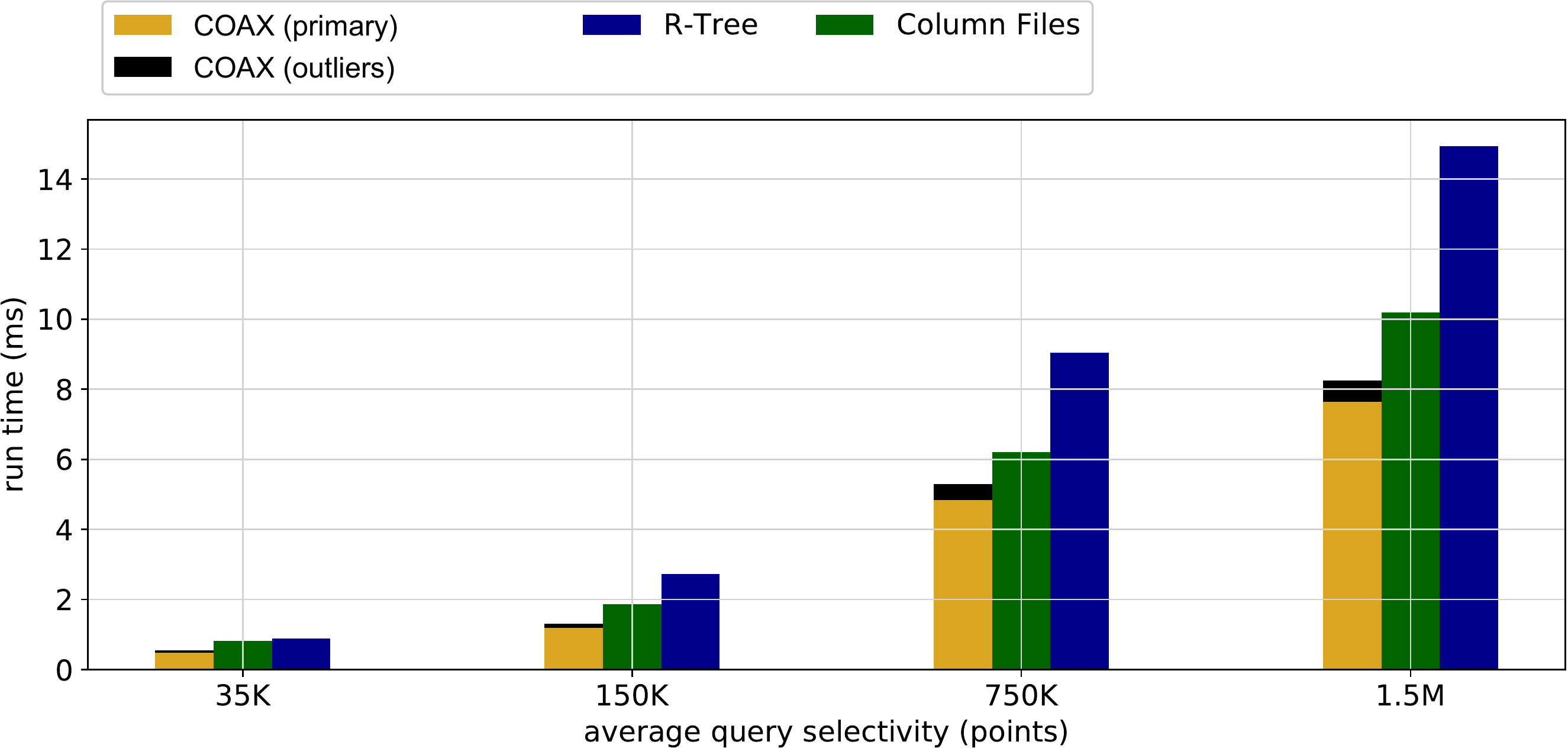}
\vspace{-7pt}
\caption{Runtime comparison for range queries with different selectivity values on airline data for year 2008 (7M)}
\label{runtimeselectivity}
\end{figure}

\subsubsection{Effect of selectivity in range queries}
In addition to this, we run the same experiment with the range queries with selectivity sizes. In this experiment we use the airline data for the year 2008 only. The results, shown in Figure~\ref{runtimeselectivity}, indicate that the Learned Gird does not lose performance on larger/shorter queries. We can check whether the query intersects with the primary, the outlier, or both indexes; and run it against the appropriate indexes. For queries with larger selectivities, it is hence more likely for both indexes to be invoked which results in more invocation and bigger overhead for the outlier index.

\todo[inline]{difference in pre processing step. and limitations}
\todo[inline]{bar plot of memory}
\todo[inline]{why the osm data doesnt perform that good}
\todo[inline]{explain what is seen in the Figures more}
\todo[inline] {make a table for run times}
\todo[inline] {normalise run times}

\subsubsection{Memory Usage}
\label{sec:memoryperformancetradeoff}

\begin{figure}
  \centering
    \begin{minipage}[b]{\linewidth}
  \centering
    \includegraphics[width=\linewidth]{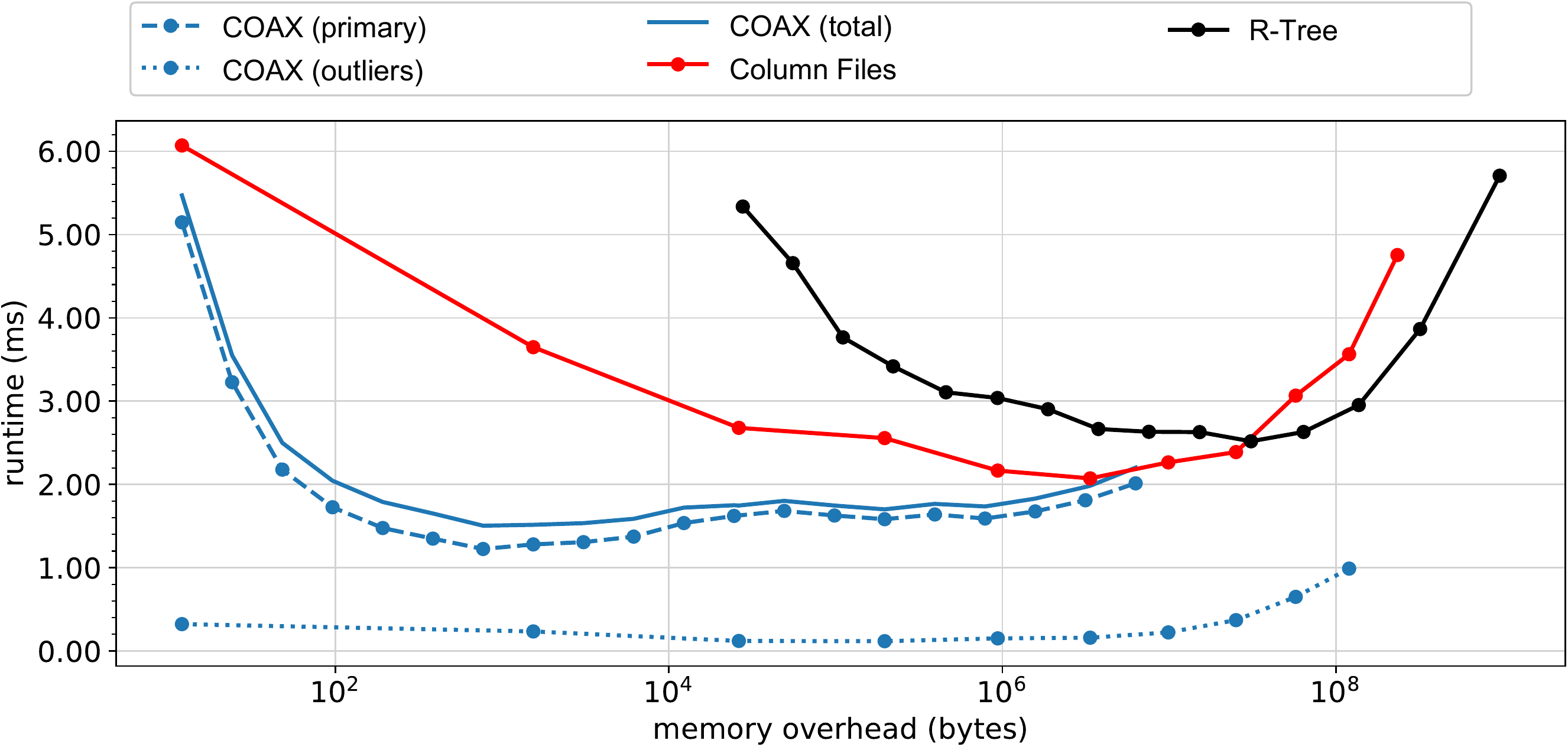}
    \subcaption{Airlines (7M)}
  \end{minipage}
  \hfill
  \centering
    \begin{minipage}[b]{\linewidth}
  \centering
    \includegraphics[width=\linewidth]{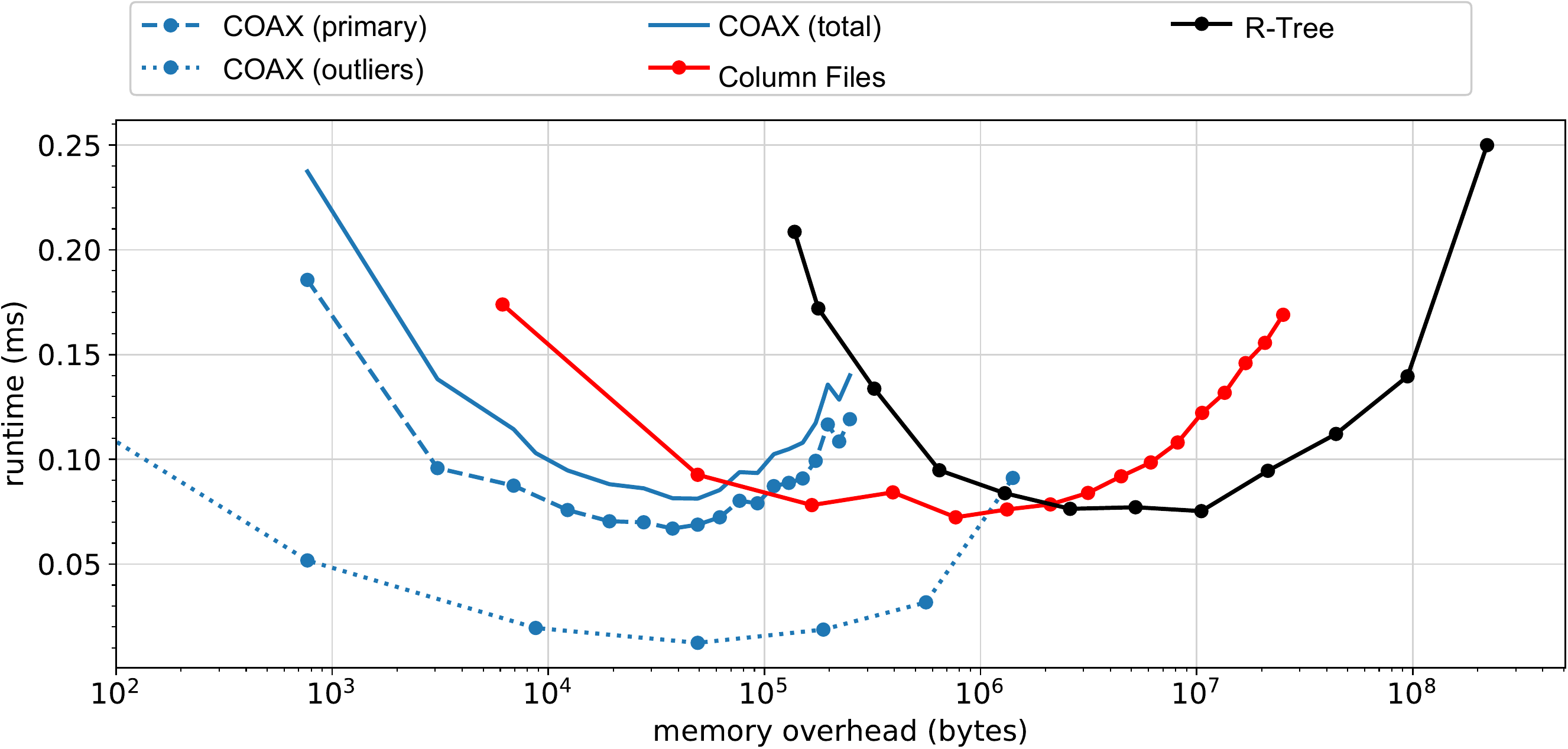}
    \subcaption{OSM (9M)}
  \end{minipage}
  \caption{Runtime vs memory footprint tradeoff for COAX and competitors. Note the log scale}
  \label{fig:pointercost}
\end{figure}

Figure~\ref{fig:pointercost} plot the range query performance against memory overhead in the case of COAX, attribute files and the R-Tree. The results show that all grid indexes have a sweet spot. This is because as we increase the number of cells, although we are likely to scan fewer items, after a certain point the increased pointer lookups starts to hurt performance mainly because there is diminishing returns in reducing the actual items considered when increasing the grid size. %
\ifextended
This effect is illustrated in Figure~\ref{gridshape}.
\fi

\section{Conclusions \& Future Work}
\label{sec:conclusions}
In this paper we address the degradation of performance/query execution time for every additionally indexed dimension. Instead of indexing all dimensions, COAX learns the soft-FDs, i.e., the correlation between different attributes. By doing so, we no longer have to index the \textit{dependent} attributes and can thus reduce the number of dimensions indexed, thereby accelerating query execution and reducing memory overhead. In case the dependent attribute is queried for, we use the model (as well as the outlier index) to find a starting point for scanning the data. As we show experimentally, our approach uses substantially less memory and also executes queries faster. COAX shrinks the memory footprint of the index by four orders of magnitude, while reducing the query execution time by 25\%.  

Our idea can be extended to the case of non-linear models for the dependency and the margins, including a mixture of models. Also, COAX can be extended to support updates. We leave this as future work.


\ifextended
\appendix

\section{Appendix Overview}

In the appendix we provide a model that theoretically proves the capacity of linear dependencies in the Soft-FD model through the approximation of new sequences and stochastic analysis. Then, we include proofs for the theorems in Section~\ref{sec:theoretical_analysis} using the suggested theoretical model. We also include a comparison between our soft-FD index and Grid search.

\section{The Center-Sequence Model (CSM)}
\label{s7_3}
In this section, we compute the space cost of linear dependencies and the capacity of the soft-FD to efficiently model complex functional dependencies in data. We focus on linear soft-FD models and determine 1) what is the capacity of a linear model for modeling soft FDs with a specific maximum error (margin), and 2) if a single linear model cannot handle the data efficiently, how many segments are needed to model the soft FDs across the entire range of the primary attribute (i.e., across $X_{range}$ in Figure~\ref{fig:theory_softfd_scan_areas}). Following the theoretical supports of the effect of the learned index, we were also inspired to make a theoretical evaluation of the capacity and the performance of the Soft-FD model. Earlier studies \cite{fotso2019grasp,guo2010improved} suggest a method of constructing a new sequence by mean values or centers is widely used in the approximation of the original data set, since the two have the same statistical properties but the new sequence ignores some noise. Therefore, combined with the similar stochastic analysis method used in \cite{ferragina2020learned}, we aim to construct a new sequence to give a  theoretical support for the effectiveness of the Soft-FD model.

We first model the dataset as a sequence of intervals using the proposed \textit{Center-Sequence Model} (CSM). Then, we perform stochastic analysis on the CSM-based representation of data to estimate the expected number of keys covered by a linear segment and  the number of segments needed to cover a sequence of n multidimensional records.

\subsection{Representing Data with CSM}
Now we suggest the Center-Sequence Model, which represents the data as a sequence containing intervals where the data points are expected to be in the centre of each interval. 
The new representation approximates the original data, such that the data and its interval-based representation have similar statistical characteristics.
The CSM model is inspired by a recent study~\cite{ferragina2020learned} on learned index structures which proposed using time series approximation and stochastic analysis to compute the upper and lower limits to estimate the capacity and the cost of using the piece-wise linear approximation (PLA) to fit the data points to a model in PLA-compatible learned index structure such as the PGM-index~\cite{ferragina2020pgm}.
The CSM representation makes it more straightforward to exploit stochastic analysis methods that we use for estimating the space cost of the soft-FD index and the effectiveness of linear and spline Soft-FD models. 
In fact, approximating data using sequences (or sub-sequences, in case of linear spline soft-FD indexes) is very common in related work \cite{fotso2019grasp,ferragina2020learned}. Based on a stochastic analysis we can give a very good estimation of the effectiveness of linear correlations and help us simplify our model. 

\subsection{Constructing new sequences to approximate the original data sequences}

We  explain how the interval representation is built from the  original data. We consider the two-dimensional case and assume that the data set is relatively large and contains N points. We split the B-box along the X-axis into n parts (n smaller parallelograms) using intervals of the same length $dx$. When $dx$ converges to 0, $n$ converges to $\infty$, hence we have smaller boxes where each box is expected to contain several original points.

\begin{figure}
\centering
\includegraphics[width=0.99\linewidth]{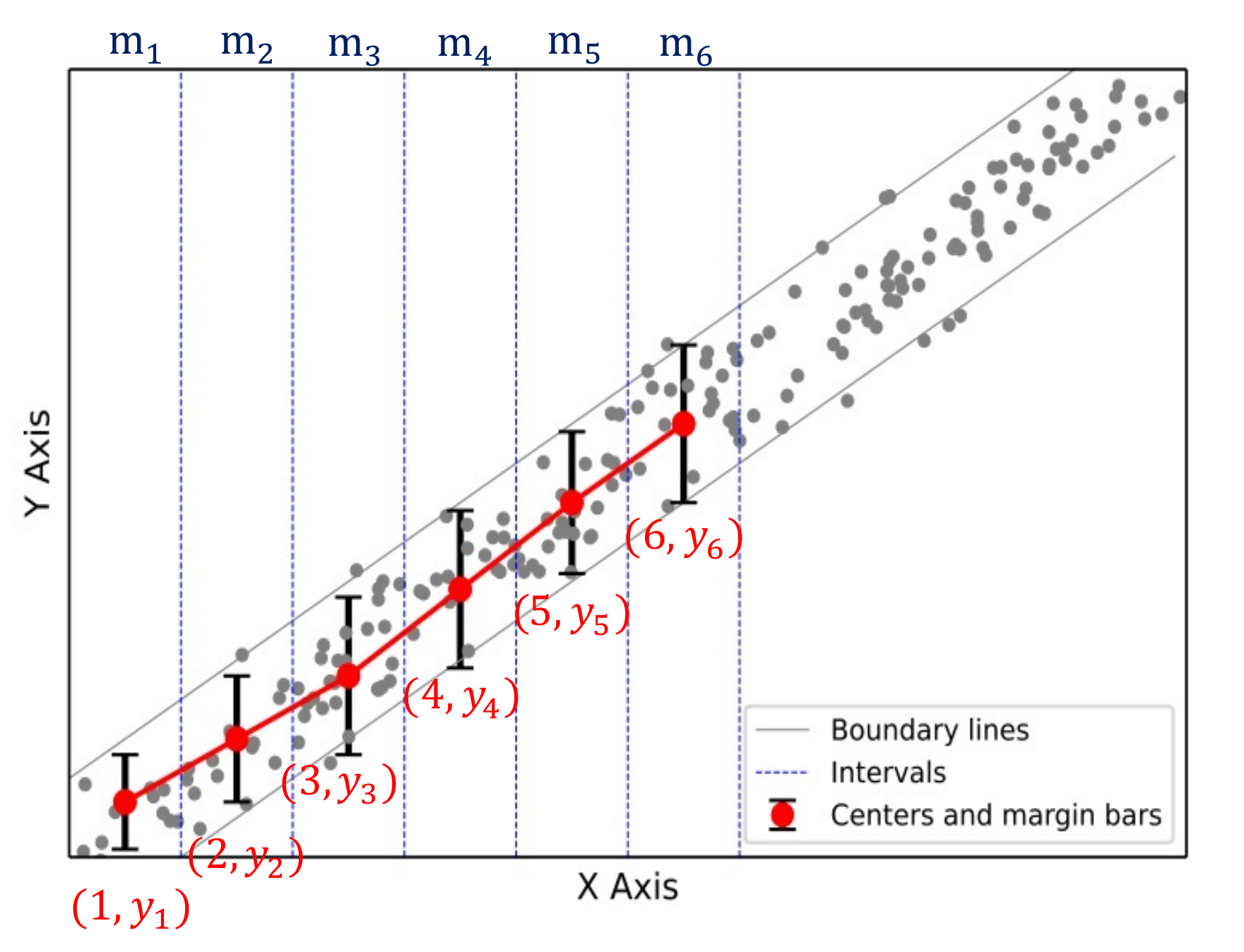}
\caption{Segmentation of the B-box using the CSM model}
\label{fig_1}
\end{figure}

Let us denote the $k^{\text{th}}$ interval as $[t_k, t_{k+1}]$, as shown in Figure~\ref{fig_1}. Within these $n$ separated parts of B-box, if we connect the centroids of these sequences to form an equally spaced sequence, the new sequence will be ${ \{ [(t_i+t_{i+1}) /2, y_{i}], i \in [1,N]\} }$, where $y_i$ is the mean value of all points in the $i^{\text{th}}$ segment along the Y axis. Let us denote the number of points located in the $k^{\text{th}}$ part as $n_i$, and the Y values of these original data points as $Y_{original}$. 

\begin{equation}
    y_i =\frac{1}{n_i} \sum_{Y_{original} \in i^{\text{th}}\,\text{interval}} Y_{original}
\end{equation}

We can rewrite the new sequence as $\{(i, y_i), i \in N\}$. In order to make our subsequent analysis more convenient, the starting point of the entire sequence is set as (0,0), i.e. when $i=0$, $y_i=0$.

For a precise CSM model, the original data points must be distributed around the centers. If we know the distribution in advance, we can precisely embed the distribution in the model. However, in a general case where we cannot make any assumptions about data distribution, the number of intervals n should be big so that the original data can be precisely approximated by the CSM model. We use the midpoints of the intervals along the X axis, say $[m_i, m_{i+1}]$,  to create equally spaced sequences. %

\subsection{Prerequisite of the efficiency of the model} 
The CSM model assumes that the data distribution on the primary attribute (X-axis) is not extremely skewed and is close to uniform. Also, it assumes that the original data is large enough such that it can be accurately approximated by the interval model. To measure how close the data distribution is to a uniform distribution, we use the Kullback-Leibler (KL) Divergence. In our case, let's denote the set of X coordinates of all points is $X_t$, and then we need to define a new set to store all unique values in $X_t$ that are not repeated, and we call it the unique set of $X_t$, denoted as $X_u$, and the number of unique values in the unique set $X_u$ is $N_{unique}$. Since we have $P_{\text{uniform},i} = 1/N_{unique},  \forall i \in N$, and given $i\in [1,N_{unique}]$,  $X_i \in X_u$, $P_i = count(X_k = X_i)/ N_i, \forall X_k \in X_t$, KL-divergence could be expressed as 
\begin{equation}
    D_{KL}(P || P_{\text{uniform}}) = \sum_i^{N_{unique}} P_i log(\frac{P_i}{P_{\text{uniform},i}})
\end{equation}

KL-Divergence provides a distance between the distribution of the  original points along the X axis and a uniform distribution. Let us denote the marginal distribution function of X axis is $P(X)$ and the distribution function of a typical uniform distribution is $P_{uniform}(X)$. When $P(X)=P_{uniform}(X)$, their KL distance is 0; otherwise, the greater the difference, the greater the distance. Since our model is under a nearly uniform distribution hypothesis, we can test that when KL distance becomes smaller and closer to 0, our model will perform better.

Figure~\ref{figt2_1} shows a worst-case scenario for the CSM model to represent the original data sets. Since the  data is skewed and the X values are far from  a uniform distribution, some segments are empty, hence the CSM model will no longer be equally spaced.

\begin{figure}
\centering
\includegraphics[width=1\linewidth]{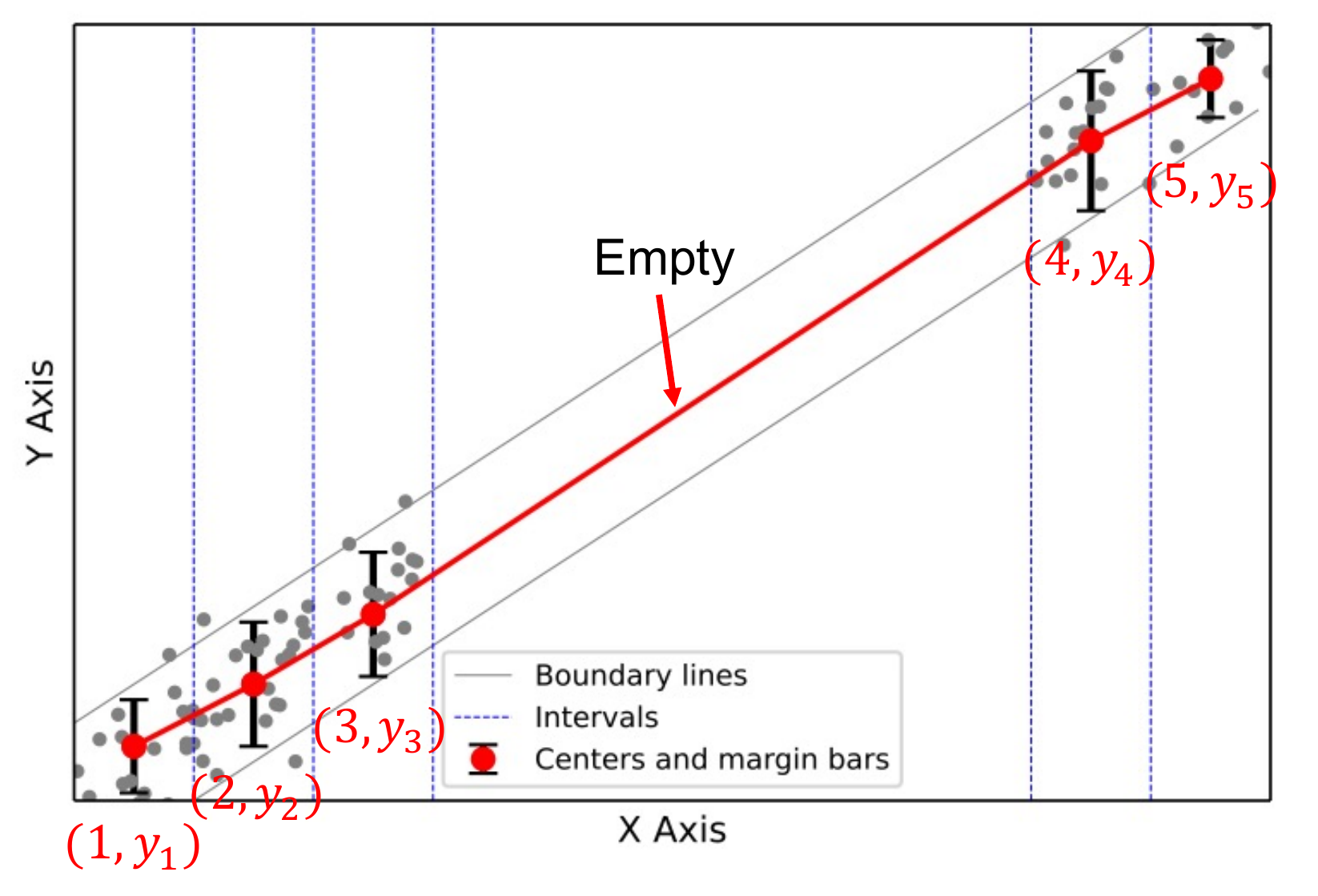}
\caption{Skewed data cases}
\label{figt2_1}
\end{figure}

\subsection{Relationship between new sequence and original one.} The goal is to show how far the original data points are from these centers. So let us assume a distribution model, i.e., we can assume all original points follow the distribution $N(y_i , \sigma_i)$, where the $y_i$ is the corresponding centers of the original points. The idea is equivalent to making assumptions on the generation rules of the original sequence given the new sequence. Then we can directly define the noise bar/ margin bar of these centers and say that all points are limited with high possibilities in a certain range around their centers. Put formally, 

\begin{equation}
\forall \  X_{original} \in [m_i,m_{i+1}], P [Y_{original} \in [y_i-  2\sigma_i, y_i +2 \sigma_i] ] = 95\%    
\end{equation}

Let us define $bar_m = max_{i} (\sigma_i), i \in N$, then we can say it is very possible (p = 95\%) that almost all original data points are limited in the $[ax+\varepsilon+2bar_m, ax+\varepsilon- 2bar_m ]$ before the fist Exit Time of new sequence {(i, $y_i$), i $\in$ N}

\begin{figure}
  \centering
  \begin{minipage}[b]{0.9\linewidth}
    \includegraphics[width=\linewidth]{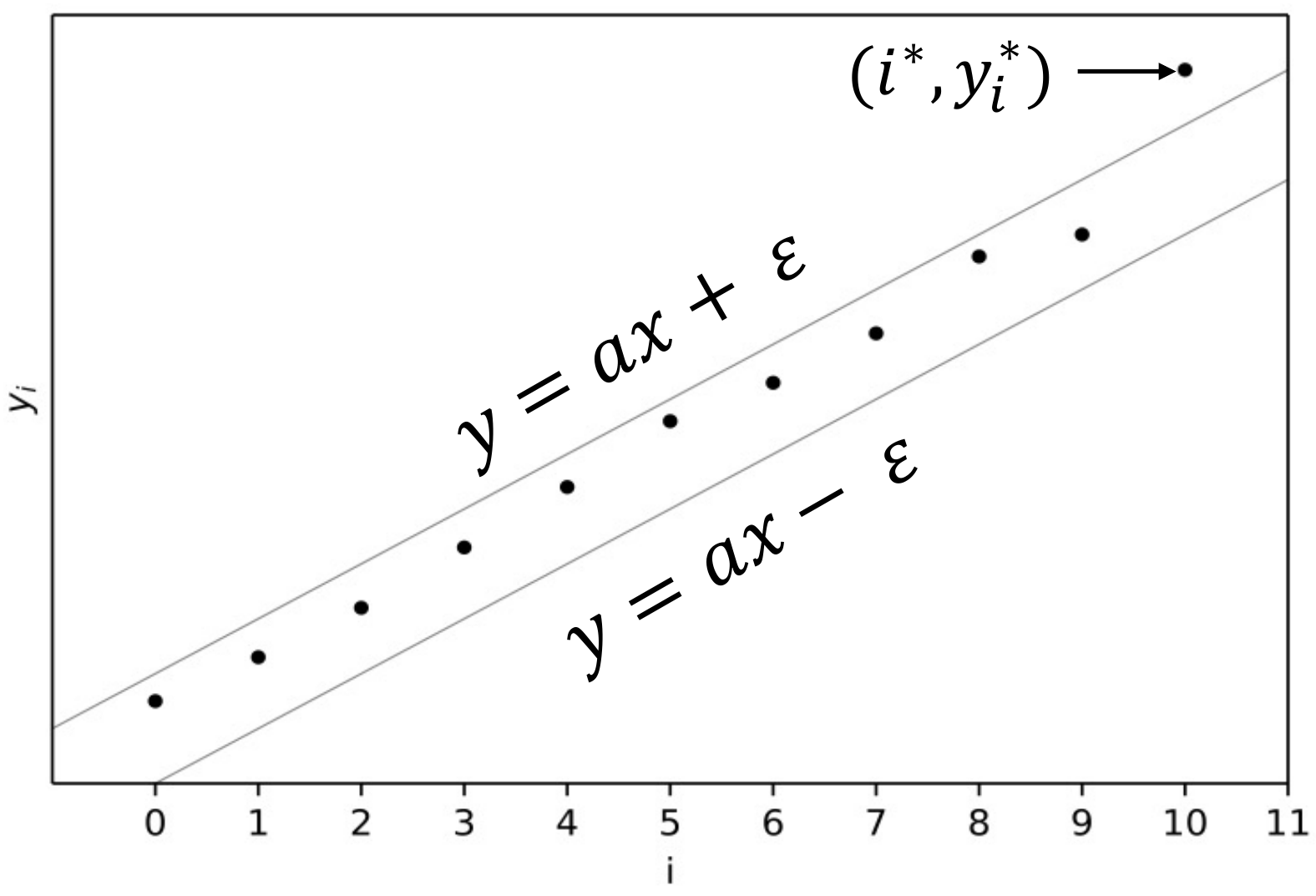}
    \subcaption{An example random walk}
  \end{minipage}
  \hfill
  \begin{minipage}[b]{0.9\linewidth}
    \includegraphics[width=\linewidth]{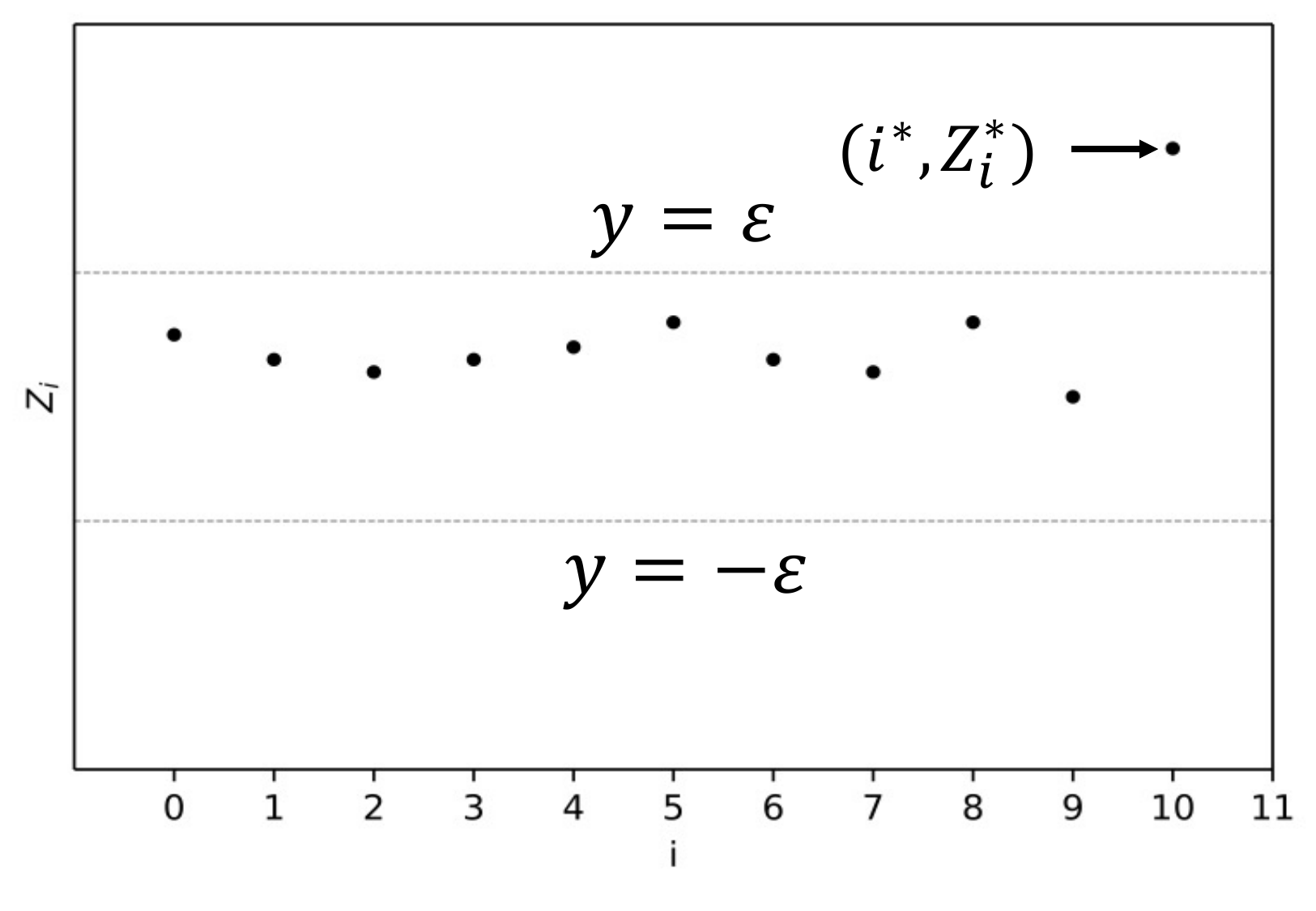}
    \subcaption{The corresponding transformed random walk}
  \end{minipage}
  \hfill
  \caption{Random walk illustration. The specific sequence model here refers to the key-position sequence proposed by Ferragina et al.~\cite{ferragina2020learned}. The difference between our model and the model in \cite{ferragina2020learned} is that for Soft-FDs we do not need to exchange the coordinate plane of the new stochastic sequence constructed by the CSM model, and the model is applicable for any unskewed value-value sequences. (a) shows the sequence of the centroids of the intervals, which are estimated using the CSM method mentioned in section \ref{s7_3}. We treat this sequence as a random walk. (b) shows the transformed random walk mentioned in section \ref{subsec:theory-capacity}, which is used in the stochastic analysis.} %
 \label{fig:transformed_random_walk}
\end{figure}

\section{Proof of Theorem~\ref{t1}}

\begin{proof}
Based on the CSM model and the reasonable assumptions made, our data points should be limited by linear segments, each of which consists of two boundary lines. Given the i.i.d. increments, the new sequence should be described as a plain stochastic process. When considering the expected numbers of keys (We call the values along Y-axis of the new sequence as "keys") covered by a linear segment, we only need to find the Mean First Exit Time (MFET) of the new sequence $\{(i, y_i), i \in N\}$.

As mentioned in \cite{gardiner1985handbook}, the Mean First Exit Time is defined from the spatial domain $D$ as $i^* := inf \{ i \geq 0, X_i \notin D \}$. Thus, in our case, the MFET can be written as $i^* = inf \{ i \in N | y_i > ai + \varepsilon \vee  y_i  <  ai - \varepsilon \}$

To make our formula more intuitive and concise, we use new symbols to simplify the stochastic sequence using the approach suggested by Ferragina et al.~\cite{ferragina2020learned}, where we consider $y_i  = \sum_{j=1}^i G_j$, and define $W_j = G_j - a$. Thus, we have $Z_i = y_i - a \cdot i = \sum_{j=1}^i (G_j - a) = \sum_{j=1}^i W_j$.

The statistical properties of the transformed stochastic process $Z_i$ and its increments $W_i$ can be easily computed based on  Ferragina et al.~\cite{ferragina2020learned}, Also, following the Central Limit Theorem, equation 3.1.(3)~\cite{ferragina2020learned}, the conditions mentioned in Masoliver et al.~\cite{masoliver2005extreme} and Redner~\cite{redner2001guide}, we know that when $\sigma \ll \varepsilon $,  the transformed stochastic process converges to a Brownian motion in continuous time.

According to \cite{gardiner1985handbook}, for a driftless Brownian motion, the Mean First Exit Time (MFET) out of an interval $[-\delta/2, \delta/2]$ is $T(x_0) = \frac{(\delta/2)^2-x_0^2}{\sigma^2}$, where $x_0$ denotes the initial position.

In our case, $x_0 =0$ and $\delta = 2\varepsilon$, thus we can easily conclude that $MFET = \frac{\varepsilon^2}{\sigma^2}$. The result means that for the new sequence, the expected numbers of keys covered by a linear segment is $MFET = \frac{\varepsilon^2}{\sigma^2}$. As we mentioned above, it is also reasonable to use this result to estimate the expected number of data points covered in the original Soft-FD model.

\end{proof}

\section{Proof of Theorem~\ref{t2}}

\begin{proof}

Following the proof of Theorem \ref{t1}, we want to express the expected number of keys covered by a single linear segment as a function of margin parameter $\varepsilon$ and the variance of the gap distribution $\sigma$. It can be easily shown that the transformed stochastic process $Z_i$ has increments with mean $d=\mu-a$ and variance $\sigma^2$. Based on prior analyses~\cite{ferragina2020learned}, when introducing the drift coefficient, the MFET out of an interval $[-\delta/2,\delta/2]$ for a Brownian motion with drift coefficient $d\rightarrow 0 $ and diffusion rate $\sigma$ can be written as  
\begin{equation}
T(0) = \frac{\varepsilon}{d} tanh(\frac{\varepsilon d}{\sigma^2})
\end{equation}
When we introduce our model and the slope $a$, it is easy to see that the maximum of $T(0)$ is achieved for $d=0$, i.e., when $a=\mu$.

\end{proof}

\section{Proof of Theorem~\ref{t3}}

\begin{proof}
Now that we have the expected value of the Mean First Exit time, we need to know the second moment of MFET to compute the variance. Following Gardiner~\cite{gardiner1985handbook}, Equation 5.2.140-156), the second moment $T_2(x)$ of the exit time of a Brownian motion with diffusion rate $\sigma$ starting at x is the solution of the partial differential equation $-2T(x) = \frac{\sigma^2}{2} \frac{\partial^2 T_2(x)}{\partial x^2}$.

Where $T(x) = \frac{(\delta/2)^2-x^2}{\sigma^2}$ is the MFET. Thus, we have 
\begin{equation}
    T_2(x) = \int -\frac{4}{\sigma^2} T(x) dx
\end{equation}

with boundary conditions $T_2(\pm\delta/2) =0$. By solving for $T_2(x)$, the second moment of the exit time is $T_2(0) = \frac{5\varepsilon^4}{3\sigma^4}$ when setting $x_0 =0 $ and $\delta = 2\varepsilon$. Therefore,

\begin{equation}
Variance = T_2(0) - [T(0)]^2 =\frac{2\varepsilon^4}{3 \sigma^4}
\end{equation}

\end{proof}

\section{Proof of Theorem~\ref{t4}}

\begin{proof}

Consider that one single linear segment cannot cover all of the data, so we can assume the total number of segments needed to be s given a stream of stochastic sequence with length n. Suggested by Ferragina et al.~\cite{ferragina2020learned}, we similarly define 
\begin{equation}
s(n) := max \{k| k \geq 1, s.t.  S_k \leq n\}
\end{equation}
where $S_k = i_1^* + i_2^*+ ... + i^*_k$.

Recall that a renewal process is an arrival process in which the interarrival intervals are positive, independent and identically distributed (i.i.d.) random variables, we notice that $\{s(n)\}_{n\geq 0}$ is a \textit{renewal counting process}. Therefore, based on~\cite{embrechts2013modelling}, $\mathbb{E}[s(n)] = \lambda n + O(1)$ as $n \rightarrow \infty$, $Var[s(n)] = \chi^2\lambda^3n + O(n)$ as $n \rightarrow \infty$, and $s(n)/n \rightarrow \lambda$. In our case, it holds: 
\begin{equation}
\frac{1}{\lambda} = \frac{\varepsilon^2}{\sigma^2} \ \  \text{and} \ \ \chi^2 =\frac{2\varepsilon^4}{3 \sigma^4}
\end{equation}
 
Hence, almost surely  $s(n)/n \rightarrow \lambda = \frac{\sigma^2}{\varepsilon^2}$. Thus, we also have $s(n) \rightarrow n \cdot \frac{\sigma^2}{\varepsilon^2}$
Given this result, we conclude that the number of segments s converges to $O(n/\varepsilon^2)$.

\end{proof}

\section{Comparison with the Grid Index}
To understand the effect of the proposed index on memory footprint reduction, we compare it with a multidimensional index, specifically a grid index. For the sake of simplicity, we compare the Soft-FD index against a square grid and do not consider workload-aware or distribution-based optimizations on grids.

Consider a grid index with N dimensions. If the data has soft functional dependencies between some dimensions, we can partition N dimensions into \textit{primary} and \textit{soft-dependent} attributes where for each attribute in the soft-dependent set, say Y, there is an attribute from the primary dimensions, say X, such that $X\rightarrow Y$ is a soft FD. Let us denote $k^{\text{th}}$ the ($k^{\text{th}}$ pair of X-Y soft functional dependency. $a_k$ is the slope of the fitted curve of Soft-FD model. 

A grid index, on the other hand, creates an N-dimensional array of cells where each cell stores points for specific ranges in each dimension. Figure~\ref{fig:square_grid_search} shows how a 2D grid processes the range query ${x \in [l_x, h_x], y \in [l_y, h_y]}$. The grid scans 14 cells for processing the query (empty cells still require a lookup). The soft FD index, however, translates the Y-axis constraints $y \in [l_y, h_y]$ to X-axis constraints $x \in [l'_x, h'_x]$ and the final constrains on the X-axis is the intersection of the two ranges ($[l'_x, h'_x] \cap [l'_x, h'_x])$. In the example shown in Figure~\ref{fig:square_grid_search}, the X-axis range inferred from the Y-axis, $[l'_x, h'_x]$, is a smaller range than the actual constraints on $X$ which is $[l_x, h_x]$. Therefore, the soft-FD index scans a smaller area ($S_s$) than the grid ($S_{\text{Grid}}$). In general, $S_s$ may be smaller or bigger than $S_{\text{Grid}}$ depending on the query constraints, the size of the grid cells, and the margins of the soft-FD index.

We analyze how reducing the dimensions affects the index structures.  We are interested to see "how many cells does the 2-D grid need to scan, compared to a soft-FD index that keeps the records in a contiguous array, assuming that both indexes scan areas of roughly the same size". Therefore, we consider a grid index in which the number of cells is specifically chosen such that the area the grid scans equals a constant value t times the scanned area in the Soft-FD model, i.e., $S_{Grid} = t \cdot S_s$. When t=1, it means we have exactly the same scanned area between the Soft-FD and grid models.

The number of cells that a square grid index scans can be estimated as the ratio between the whole area and the scanned area.

\begin{equation}
    n_k =\frac{S_{whole}}{t \cdot S_s}  =  (\frac{Y_{range}}{t \cdot \varepsilon} \cdot \frac{X_{range}}{(2\varepsilon+q_y)/a_k})
\end{equation}

\begin{figure}
\centering
\includegraphics[width=0.99\linewidth]{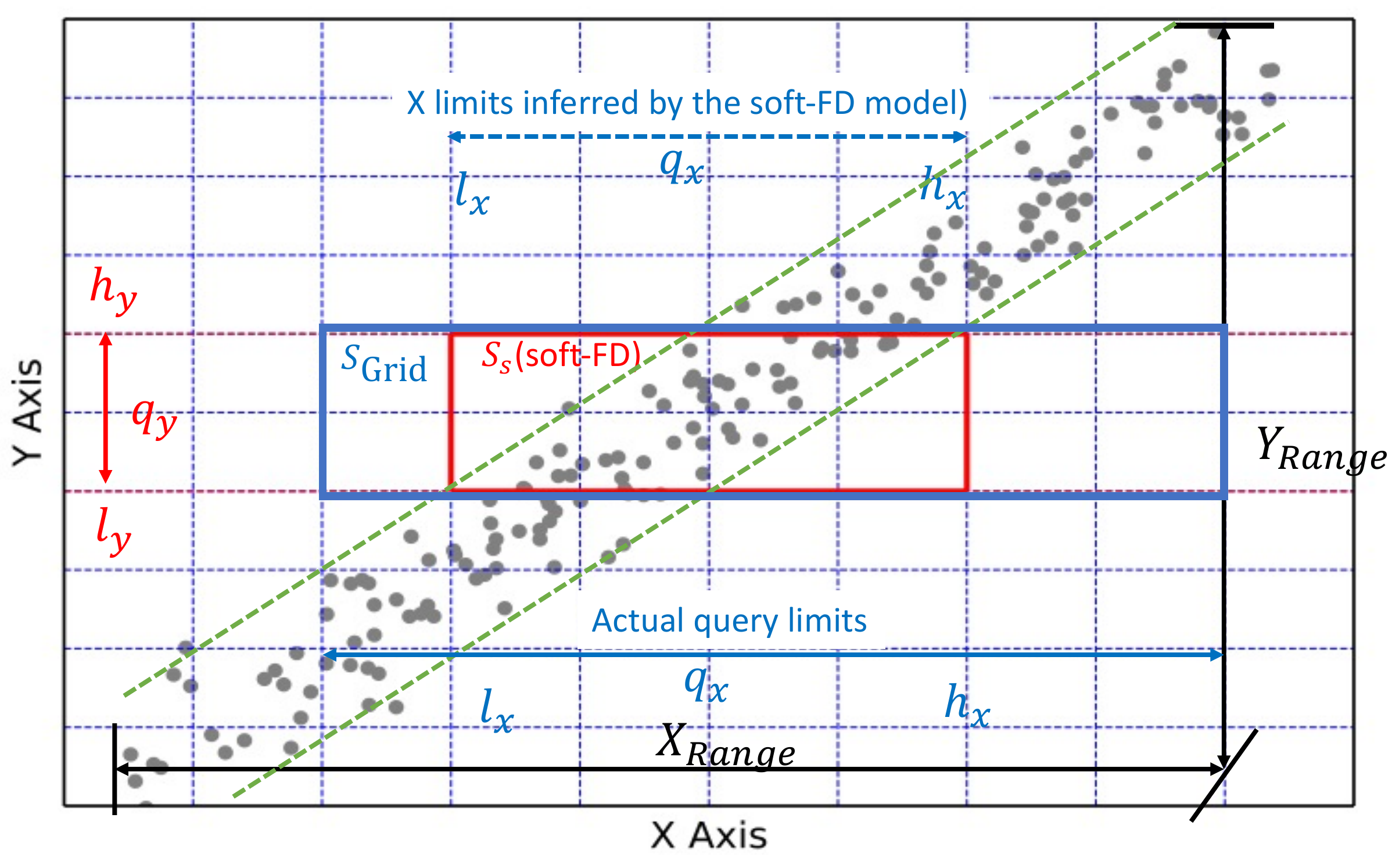}
\vspace{-8pt}
\caption{An example Square-Grid.  The area that the grid scans for the range query $\{ (x,y) | l_x \leq x \leq h_x, l_y \leq y \leq h_y)\}$ is marked with a red rectangle.}
\label{fig:square_grid_search}
\end{figure}

We define $r_k$ as the ratio between length and the width of s-box:
\begin{equation}
    r_k = \frac{\text{Length(B-box)}}{\text{Width(B-Box)}} = \frac{\sqrt{Y_{range}^2+X_{range}^2}}{2\varepsilon / \sqrt{1+a_k^2}}
\end{equation}

Then, we can compute the scanned cells in grid as:

\begin{equation*}
\begin{aligned}
n_k &= \frac{X_{range} \cdot Y_{Range} \cdot a_k}{t \cdot ( \frac{(1+a_k^2)(Y_{range}^2+X_{range}^2)}{r_k^2} + q_y \frac{(\sqrt{1+a_k^2})(\sqrt{Y_{range}^2+X_{range}^2}}{r_k} ) } \\
    &= \frac{X_{range} \cdot Y_{Range} \cdot a_k}{t \cdot (f(r_k|a_k,X,Y)^2+ q_y f(r_k|a_k,X,Y))}
\end{aligned}
\end{equation*}

Where $f(r_k|a_k,X,Y)=(\sqrt{1+a_k^2})(\sqrt{Y_{range}^2+X_{range}^2})/r_k $ is a function of $r_k$. Given the distribution of data points and a fixed linear dependence, when $r_k$ increases,  $f(r_k|a_k,X,Y)$ decreases, and then $n_k$ increases. It can be easily imagined that when $ r_k \rightarrow \infty$, $n_k$ will get close to infinity as well. Thus, for K pairs of attributes in total (K is O(d), d denotes the total dimensions), we may need to create $n_1\cdot n_2 ... \cdot n_K = n^K$ cells in total. It can be easily concluded that if $r_k$ is larger, i.e., the B-box has a narrow margin, therefore and equivalent grid index needs a large number of cells to reach the almost same space cost as Soft-FD model.

\fi


\begin{thebibliography}{37}

%%% the Web address.

\ifx \showCODEN    \undefined \def \showCODEN     #1{\unskip}     \fi
\ifx \showDOI      \undefined \def \showDOI       #1{#1}\fi
\ifx \showISBNx    \undefined \def \showISBNx     #1{\unskip}     \fi
\ifx \showISBNxiii \undefined \def \showISBNxiii  #1{\unskip}     \fi
\ifx \showISSN     \undefined \def \showISSN      #1{\unskip}     \fi
\ifx \showLCCN     \undefined \def \showLCCN      #1{\unskip}     \fi
\ifx \shownote     \undefined \def \shownote      #1{#1}          \fi
\ifx \showarticletitle \undefined \def \showarticletitle #1{#1}   \fi
\ifx \showURL      \undefined \def \showURL       {\relax}        \fi
% invisible to TeX
\providecommand\bibfield[2]{#2}
\providecommand\bibinfo[2]{#2}
\providecommand\natexlab[1]{#1}
\providecommand\showeprint[2][]{arXiv:#2}

\bibitem[\protect\citeauthoryear{??}{osm}{}]%
        {osm}
 \bibinfo{year}{[n.d.]}\natexlab{}.
\newblock \bibinfo{title}{OpenStreetMap Data}.
\newblock \bibinfo{howpublished}{\url{https://download.geofabrik.de}}.
\newblock
\newblock
\shownote{2020-12-22.}


\bibitem[\protect\citeauthoryear{??}{pym}{}]%
        {pymc3}
 \bibinfo{year}{[n.d.]}\natexlab{}.
\newblock \bibinfo{title}{Probabilistic Programming in Python: Bayesian
  Modeling and Probabilistic Machine Learning with Theano}.
\newblock \bibinfo{howpublished}{\url{https://github.com/pymc-devs/pymc3}}.
\newblock
\newblock
\shownote{2020-12-22.}


\bibitem[\protect\citeauthoryear{Ao, Zhang, Wu, Stones, Wang, Liu, Liu, and
  Lin}{Ao et~al\mbox{.}}{2011}]%
        {ao2011efficient}
\bibfield{author}{\bibinfo{person}{Naiyong Ao}, \bibinfo{person}{Fan Zhang},
  \bibinfo{person}{Di Wu}, \bibinfo{person}{Douglas~S Stones},
  \bibinfo{person}{Gang Wang}, \bibinfo{person}{Xiaoguang Liu},
  \bibinfo{person}{Jing Liu}, {and} \bibinfo{person}{Sheng Lin}.}
  \bibinfo{year}{2011}\natexlab{}.
\newblock \showarticletitle{Efficient parallel lists intersection and index
  compression algorithms using graphics processing units}.
\newblock \bibinfo{journal}{\emph{PVLDB}} \bibinfo{volume}{4},
  \bibinfo{number}{8} (\bibinfo{year}{2011}), \bibinfo{pages}{470--481}.
\newblock


\bibitem[\protect\citeauthoryear{Dai and Shrivastava}{Dai and
  Shrivastava}{2019}]%
        {dai2019adaptive}
\bibfield{author}{\bibinfo{person}{Zhenwei Dai} {and}
  \bibinfo{person}{Anshumali Shrivastava}.} \bibinfo{year}{2019}\natexlab{}.
\newblock \showarticletitle{Adaptive learned Bloom filter (Ada-BF): Efficient
  utilization of the classifier}.
\newblock \bibinfo{journal}{\emph{arXiv:1910.09131}} (\bibinfo{year}{2019}).
\newblock


\bibitem[\protect\citeauthoryear{Ding, Minhas, Yu, Wang, Do, Li, Zhang,
  Chandramouli, Gehrke, and Kossmann}{Ding et~al\mbox{.}}{2020}]%
        {ding2020alex}
\bibfield{author}{\bibinfo{person}{Jialin Ding}, \bibinfo{person}{Umar~Farooq
  Minhas}, \bibinfo{person}{Jia Yu}, \bibinfo{person}{Chi Wang},
  \bibinfo{person}{Jaeyoung Do}, \bibinfo{person}{Yinan Li},
  \bibinfo{person}{Hantian Zhang}, \bibinfo{person}{Badrish Chandramouli},
  \bibinfo{person}{Johannes Gehrke}, {and} \bibinfo{person}{Donald Kossmann}.}
  \bibinfo{year}{2020}\natexlab{}.
\newblock \showarticletitle{ALEX: an updatable adaptive learned index}. In
  \bibinfo{booktitle}{\emph{SIGMOD}}.
\newblock


\bibitem[\protect\citeauthoryear{Embrechts, Kl{\"u}ppelberg, and
  Mikosch}{Embrechts et~al\mbox{.}}{2013}]%
        {embrechts2013modelling}
\bibfield{author}{\bibinfo{person}{Paul Embrechts}, \bibinfo{person}{Claudia
  Kl{\"u}ppelberg}, {and} \bibinfo{person}{Thomas Mikosch}.}
  \bibinfo{year}{2013}\natexlab{}.
\newblock \bibinfo{booktitle}{\emph{Modelling extremal events: for insurance
  and finance}}. Vol.~\bibinfo{volume}{33}.
\newblock \bibinfo{publisher}{Springer}.
\newblock


\bibitem[\protect\citeauthoryear{Ferragina, Lillo, and Vinciguerra}{Ferragina
  et~al\mbox{.}}{2020}]%
        {ferragina2020learned}
\bibfield{author}{\bibinfo{person}{Paolo Ferragina}, \bibinfo{person}{Fabrizio
  Lillo}, {and} \bibinfo{person}{Giorgio Vinciguerra}.}
  \bibinfo{year}{2020}\natexlab{}.
\newblock \showarticletitle{Why are learned indexes so effective?}. In
  \bibinfo{booktitle}{\emph{ICML}}.
\newblock


\bibitem[\protect\citeauthoryear{Ferragina and Vinciguerra}{Ferragina and
  Vinciguerra}{2020a}]%
        {ferragina2020survey}
\bibfield{author}{\bibinfo{person}{Paolo Ferragina} {and}
  \bibinfo{person}{Giorgio Vinciguerra}.} \bibinfo{year}{2020}\natexlab{a}.
\newblock \showarticletitle{Learned data structures}.
\newblock \bibinfo{journal}{\emph{Recent Trends in Learning From Data}}
  (\bibinfo{year}{2020}), \bibinfo{pages}{5--41}.
\newblock


\bibitem[\protect\citeauthoryear{Ferragina and Vinciguerra}{Ferragina and
  Vinciguerra}{2020b}]%
        {ferragina2020pgm}
\bibfield{author}{\bibinfo{person}{Paolo Ferragina} {and}
  \bibinfo{person}{Giorgio Vinciguerra}.} \bibinfo{year}{2020}\natexlab{b}.
\newblock \showarticletitle{The PGM-index: a fully-dynamic compressed learned
  index with provable worst-case bounds}.
\newblock \bibinfo{journal}{\emph{PVLDB}} \bibinfo{volume}{13},
  \bibinfo{number}{8} (\bibinfo{year}{2020}), \bibinfo{pages}{1162--1175}.
\newblock


\bibitem[\protect\citeauthoryear{Fotso, Nguifo, and Vaslin}{Fotso
  et~al\mbox{.}}{2019}]%
        {fotso2019grasp}
\bibfield{author}{\bibinfo{person}{Vanel Steve~Siyou Fotso},
  \bibinfo{person}{Engelbert~Mephu Nguifo}, {and} \bibinfo{person}{Philippe
  Vaslin}.} \bibinfo{year}{2019}\natexlab{}.
\newblock \showarticletitle{Grasp heuristic for time series compression with
  piecewise aggregate approximation}.
\newblock \bibinfo{journal}{\emph{RAIRO-Operations Research}}
  \bibinfo{volume}{53}, \bibinfo{number}{1} (\bibinfo{year}{2019}),
  \bibinfo{pages}{243--259}.
\newblock


\bibitem[\protect\citeauthoryear{Galakatos, Markovitch, Binnig, Fonseca, and
  Kraska}{Galakatos et~al\mbox{.}}{2019}]%
        {galakatos2018tree}
\bibfield{author}{\bibinfo{person}{Alex Galakatos}, \bibinfo{person}{Michael
  Markovitch}, \bibinfo{person}{Carsten Binnig}, \bibinfo{person}{Rodrigo
  Fonseca}, {and} \bibinfo{person}{Tim Kraska}.}
  \bibinfo{year}{2019}\natexlab{}.
\newblock \showarticletitle{FITing-Tree: A Data-aware Index Structure}.
\newblock  (\bibinfo{year}{2019}).
\newblock


\bibitem[\protect\citeauthoryear{Gardiner}{Gardiner}{1985}]%
        {gardiner1985handbook}
\bibfield{author}{\bibinfo{person}{Crispin~W Gardiner}.}
  \bibinfo{year}{1985}\natexlab{}.
\newblock \bibinfo{booktitle}{\emph{Handbook of stochastic methods}}.
  Vol.~\bibinfo{volume}{3}.
\newblock \bibinfo{publisher}{Springer}.
\newblock


\bibitem[\protect\citeauthoryear{Guo, Li, and Pan}{Guo et~al\mbox{.}}{2010}]%
        {guo2010improved}
\bibfield{author}{\bibinfo{person}{Chonghui Guo}, \bibinfo{person}{Hailin Li},
  {and} \bibinfo{person}{Donghua Pan}.} \bibinfo{year}{2010}\natexlab{}.
\newblock \showarticletitle{An improved piecewise aggregate approximation based
  on statistical features for time series mining}. In
  \bibinfo{booktitle}{\emph{KSEM}}. Springer, \bibinfo{pages}{234--244}.
\newblock


\bibitem[\protect\citeauthoryear{Hadian and Heinis}{Hadian and Heinis}{2019a}]%
        {hadian2019considerations}
\bibfield{author}{\bibinfo{person}{Ali Hadian} {and} \bibinfo{person}{Thomas
  Heinis}.} \bibinfo{year}{2019}\natexlab{a}.
\newblock \showarticletitle{Considerations for handling updates in learned
  index structures}. In \bibinfo{booktitle}{\emph{AIDM}}.
\newblock


\bibitem[\protect\citeauthoryear{Hadian and Heinis}{Hadian and Heinis}{2019b}]%
        {hadian2019interpolation}
\bibfield{author}{\bibinfo{person}{Ali Hadian} {and} \bibinfo{person}{Thomas
  Heinis}.} \bibinfo{year}{2019}\natexlab{b}.
\newblock \showarticletitle{Interpolation-friendly B-trees: Bridging the Gap
  Between Algorithmic and Learned Indexes}. In
  \bibinfo{booktitle}{\emph{EDBT}}.
\newblock


\bibitem[\protect\citeauthoryear{Hadian and Heinis}{Hadian and Heinis}{2020}]%
        {hadian2020madex}
\bibfield{author}{\bibinfo{person}{Ali Hadian} {and} \bibinfo{person}{Thomas
  Heinis}.} \bibinfo{year}{2020}\natexlab{}.
\newblock \showarticletitle{MADEX: Learning-augmented Algorithmic Index
  Structures}. In \bibinfo{booktitle}{\emph{AIDB}}.
\newblock


\bibitem[\protect\citeauthoryear{Hadian, Kumar, and Heinis}{Hadian
  et~al\mbox{.}}{2020}]%
        {hadian2020handsoff}
\bibfield{author}{\bibinfo{person}{Ali Hadian}, \bibinfo{person}{Ankit Kumar},
  {and} \bibinfo{person}{Thomas Heinis}.} \bibinfo{year}{2020}\natexlab{}.
\newblock \showarticletitle{Hands-off Model Integration in Spatial Index
  Structures}. In \bibinfo{booktitle}{\emph{AIDB}}.
\newblock


\bibitem[\protect\citeauthoryear{Ilyas, Markl, Haas, Brown, and
  Aboulnaga}{Ilyas et~al\mbox{.}}{2004}]%
        {ilyas2004cords}
\bibfield{author}{\bibinfo{person}{Ihab~F Ilyas}, \bibinfo{person}{Volker
  Markl}, \bibinfo{person}{Peter Haas}, \bibinfo{person}{Paul Brown}, {and}
  \bibinfo{person}{Ashraf Aboulnaga}.} \bibinfo{year}{2004}\natexlab{}.
\newblock \showarticletitle{CORDS: automatic discovery of correlations and soft
  functional dependencies}. In \bibinfo{booktitle}{\emph{SIGMOD}}.
  \bibinfo{pages}{647--658}.
\newblock


\bibitem[\protect\citeauthoryear{Kimura, Huo, Rasin, Madden, and Zdonik}{Kimura
  et~al\mbox{.}}{2009}]%
        {kimura2009correlation}
\bibfield{author}{\bibinfo{person}{Hideaki Kimura}, \bibinfo{person}{George
  Huo}, \bibinfo{person}{Alexander Rasin}, \bibinfo{person}{Samuel Madden},
  {and} \bibinfo{person}{Stanley~B Zdonik}.} \bibinfo{year}{2009}\natexlab{}.
\newblock \showarticletitle{Correlation maps: a compressed access method for
  exploiting soft functional dependencies}.
\newblock \bibinfo{journal}{\emph{PVLDB}} \bibinfo{volume}{2},
  \bibinfo{number}{1} (\bibinfo{year}{2009}), \bibinfo{pages}{1222--1233}.
\newblock


\bibitem[\protect\citeauthoryear{Kimura, Huo, Rasin, Madden, and Zdonik}{Kimura
  et~al\mbox{.}}{2010}]%
        {zdonik2010coradd}
\bibfield{author}{\bibinfo{person}{Hideaki Kimura}, \bibinfo{person}{George
  Huo}, \bibinfo{person}{Alexander Rasin}, \bibinfo{person}{Samuel~R Madden},
  {and} \bibinfo{person}{Stanley~B Zdonik}.} \bibinfo{year}{2010}\natexlab{}.
\newblock \showarticletitle{CORADD: Correlation Aware Database Designer for
  Materialized Views and Indexes}.
\newblock \bibinfo{journal}{\emph{PVLDB}} \bibinfo{volume}{3},
  \bibinfo{number}{1} (\bibinfo{year}{2010}).
\newblock


\bibitem[\protect\citeauthoryear{Kipf, Marcus, van Renen, Stoian, Kemper,
  Kraska, and Neumann}{Kipf et~al\mbox{.}}{2020}]%
        {kipf2020radixspline}
\bibfield{author}{\bibinfo{person}{Andreas Kipf}, \bibinfo{person}{Ryan
  Marcus}, \bibinfo{person}{Alexander van Renen}, \bibinfo{person}{Mihail
  Stoian}, \bibinfo{person}{Alfons Kemper}, \bibinfo{person}{Tim Kraska}, {and}
  \bibinfo{person}{Thomas Neumann}.} \bibinfo{year}{2020}\natexlab{}.
\newblock \showarticletitle{RadixSpline: A Single-Pass Learned Index}.
\newblock  (\bibinfo{year}{2020}).
\newblock


\bibitem[\protect\citeauthoryear{Kraska, Beutel, Chi, Dean, and
  Polyzotis}{Kraska et~al\mbox{.}}{2018}]%
        {kraska2018case}
\bibfield{author}{\bibinfo{person}{Tim Kraska}, \bibinfo{person}{Alex Beutel},
  \bibinfo{person}{Ed~H Chi}, \bibinfo{person}{Jeffrey Dean}, {and}
  \bibinfo{person}{Neoklis Polyzotis}.} \bibinfo{year}{2018}\natexlab{}.
\newblock \showarticletitle{The case for learned index structures}. In
  \bibinfo{booktitle}{\emph{SIGMOD}}. \bibinfo{pages}{489--504}.
\newblock


\bibitem[\protect\citeauthoryear{Li, Liang, Bai, Zheng, Xiong, and Sun}{Li
  et~al\mbox{.}}{2019}]%
        {li2019accelerating}
\bibfield{author}{\bibinfo{person}{Zhao~Lucis Li},
  \bibinfo{person}{Chieh-Jan~Mike Liang}, \bibinfo{person}{Wei Bai},
  \bibinfo{person}{Qiming Zheng}, \bibinfo{person}{Yongqiang Xiong}, {and}
  \bibinfo{person}{Guangzhong Sun}.} \bibinfo{year}{2019}\natexlab{}.
\newblock \showarticletitle{Accelerating rule-matching systems with learned
  rankers}. In \bibinfo{booktitle}{\emph{{USENIX} Annual Technical
  Conference}}. \bibinfo{pages}{1041--1048}.
\newblock


\bibitem[\protect\citeauthoryear{Liu, Xiao, Didwania, and Eltabakh}{Liu
  et~al\mbox{.}}{2016}]%
        {liu2016exploiting}
\bibfield{author}{\bibinfo{person}{Hai Liu}, \bibinfo{person}{Dongqing Xiao},
  \bibinfo{person}{Pankaj Didwania}, {and} \bibinfo{person}{Mohamed~Y
  Eltabakh}.} \bibinfo{year}{2016}\natexlab{}.
\newblock \showarticletitle{Exploiting soft and hard correlations in big data
  query optimization}.
\newblock \bibinfo{journal}{\emph{PVLDB}} \bibinfo{volume}{9},
  \bibinfo{number}{12} (\bibinfo{year}{2016}), \bibinfo{pages}{1005--1016}.
\newblock


\bibitem[\protect\citeauthoryear{Llavesh, Sirin, West, and Ailamaki}{Llavesh
  et~al\mbox{.}}{2019}]%
        {llaveshi2019accelerating}
\bibfield{author}{\bibinfo{person}{Anisa Llavesh}, \bibinfo{person}{Utku
  Sirin}, \bibinfo{person}{Robert West}, {and} \bibinfo{person}{Anastasia
  Ailamaki}.} \bibinfo{year}{2019}\natexlab{}.
\newblock \showarticletitle{Accelerating B+tree Search by Using Simple Machine
  Learning Techniques}. In \bibinfo{booktitle}{\emph{AIDB}}.
\newblock


\bibitem[\protect\citeauthoryear{Masoliver, Montero, and Perell{\'o}}{Masoliver
  et~al\mbox{.}}{2005}]%
        {masoliver2005extreme}
\bibfield{author}{\bibinfo{person}{Jaume Masoliver}, \bibinfo{person}{Miquel
  Montero}, {and} \bibinfo{person}{Josep Perell{\'o}}.}
  \bibinfo{year}{2005}\natexlab{}.
\newblock \showarticletitle{Extreme times in financial markets}.
\newblock \bibinfo{journal}{\emph{Physical Review E}} \bibinfo{volume}{71},
  \bibinfo{number}{5} (\bibinfo{year}{2005}).
\newblock


\bibitem[\protect\citeauthoryear{Mitzenmacher}{Mitzenmacher}{2018}]%
        {mitzenmacher2018model}
\bibfield{author}{\bibinfo{person}{Michael Mitzenmacher}.}
  \bibinfo{year}{2018}\natexlab{}.
\newblock \showarticletitle{A model for learned bloom filters, and optimizing
  by sandwiching}. In \bibinfo{booktitle}{\emph{NIPS}}.
  \bibinfo{pages}{462--471}.
\newblock


\bibitem[\protect\citeauthoryear{Nathan, Ding, Alizadeh, and Kraska}{Nathan
  et~al\mbox{.}}{2020}]%
        {nathan2020learning}
\bibfield{author}{\bibinfo{person}{Vikram Nathan}, \bibinfo{person}{Jialin
  Ding}, \bibinfo{person}{Mohammad Alizadeh}, {and} \bibinfo{person}{Tim
  Kraska}.} \bibinfo{year}{2020}\natexlab{}.
\newblock \showarticletitle{Learning Multi-dimensional Indexes}. In
  \bibinfo{booktitle}{\emph{SIGMOD}}. \bibinfo{pages}{985--1000}.
\newblock


\bibitem[\protect\citeauthoryear{Nievergelt, Hinterberger, and
  Sevcik}{Nievergelt et~al\mbox{.}}{1981}]%
        {nievergelt1981grid}
\bibfield{author}{\bibinfo{person}{J{\"u}rg Nievergelt}, \bibinfo{person}{Hans
  Hinterberger}, {and} \bibinfo{person}{Kenneth~C Sevcik}.}
  \bibinfo{year}{1981}\natexlab{}.
\newblock \showarticletitle{The grid file: An adaptable, symmetric multi-key
  file structure}. In \bibinfo{booktitle}{\emph{ECI}}. Springer,
  \bibinfo{pages}{236--251}.
\newblock


\bibitem[\protect\citeauthoryear{Oosterhuis, Culpepper, and
  de~Rijke}{Oosterhuis et~al\mbox{.}}{2018}]%
        {oosterhuis2018potential}
\bibfield{author}{\bibinfo{person}{Harrie Oosterhuis}, \bibinfo{person}{J~Shane
  Culpepper}, {and} \bibinfo{person}{Maarten de Rijke}.}
  \bibinfo{year}{2018}\natexlab{}.
\newblock \showarticletitle{The potential of learned index structures for index
  compression}. In \bibinfo{booktitle}{\emph{ADCS}}. \bibinfo{pages}{1--4}.
\newblock


\bibitem[\protect\citeauthoryear{Redner}{Redner}{2001}]%
        {redner2001guide}
\bibfield{author}{\bibinfo{person}{Sidney Redner}.}
  \bibinfo{year}{2001}\natexlab{}.
\newblock \bibinfo{booktitle}{\emph{A guide to first-passage processes}}.
\newblock \bibinfo{publisher}{Cambridge Uni. Press}.
\newblock


\bibitem[\protect\citeauthoryear{Setiawan, Rubinstein, and
  Borovica-Gajic}{Setiawan et~al\mbox{.}}{2020}]%
        {setiawan2020function}
\bibfield{author}{\bibinfo{person}{Naufal~Fikri Setiawan},
  \bibinfo{person}{Benjamin~IP Rubinstein}, {and} \bibinfo{person}{Renata
  Borovica-Gajic}.} \bibinfo{year}{2020}\natexlab{}.
\newblock \showarticletitle{Function Interpolation for Learned Index
  Structures}. In \bibinfo{booktitle}{\emph{ADC}}.
\newblock


\bibitem[\protect\citeauthoryear{Wang, Dong, Sarma, Franklin, and Halevy}{Wang
  et~al\mbox{.}}{2009}]%
        {wang2009functional}
\bibfield{author}{\bibinfo{person}{Daisy~Zhe Wang}, \bibinfo{person}{Xin~Luna
  Dong}, \bibinfo{person}{Anish~Das Sarma}, \bibinfo{person}{Michael~J.
  Franklin}, {and} \bibinfo{person}{Alon~Y. Halevy}.}
  \bibinfo{year}{2009}\natexlab{}.
\newblock \showarticletitle{Functional Dependency Generation and Applications
  in Pay-As-You-Go Data Integration Systems}. In
  \bibinfo{booktitle}{\emph{WEBDB}}.
\newblock


\bibitem[\protect\citeauthoryear{Wang, Tang, Wang, and Chen}{Wang
  et~al\mbox{.}}{2020}]%
        {wang2020sindex}
\bibfield{author}{\bibinfo{person}{Youyun Wang}, \bibinfo{person}{Chuzhe Tang},
  \bibinfo{person}{Zhaoguo Wang}, {and} \bibinfo{person}{Haibo Chen}.}
  \bibinfo{year}{2020}\natexlab{}.
\newblock \showarticletitle{SIndex: a scalable learned index for string keys}.
  In \bibinfo{booktitle}{\emph{{APSys}}}.
\newblock


\bibitem[\protect\citeauthoryear{Wu, Yu, Tian, Sidle, and Barber}{Wu
  et~al\mbox{.}}{2019}]%
        {wu2019designing}
\bibfield{author}{\bibinfo{person}{Yingjun Wu}, \bibinfo{person}{Jia Yu},
  \bibinfo{person}{Yuanyuan Tian}, \bibinfo{person}{Richard Sidle}, {and}
  \bibinfo{person}{Ronald Barber}.} \bibinfo{year}{2019}\natexlab{}.
\newblock \showarticletitle{Designing succinct secondary indexing mechanism by
  exploiting column correlations}. In \bibinfo{booktitle}{\emph{SIGMOD}}.
  \bibinfo{pages}{1223--1240}.
\newblock


\bibitem[\protect\citeauthoryear{Xiang, Zhang, Cui, Chu, Li, and Zhou}{Xiang
  et~al\mbox{.}}{2018}]%
        {xiang2018pavo}
\bibfield{author}{\bibinfo{person}{Wenkun Xiang}, \bibinfo{person}{Hao Zhang},
  \bibinfo{person}{Rui Cui}, \bibinfo{person}{Xing Chu}, \bibinfo{person}{Keqin
  Li}, {and} \bibinfo{person}{Wei Zhou}.} \bibinfo{year}{2018}\natexlab{}.
\newblock \showarticletitle{Pavo: A RNN-Based Learned Inverted Index,
  Supervised or Unsupervised?}
\newblock \bibinfo{journal}{\emph{IEEE Access}}  \bibinfo{volume}{7}
  (\bibinfo{year}{2018}), \bibinfo{pages}{293--303}.
\newblock


\bibitem[\protect\citeauthoryear{Zhang, Lin, and Ross}{Zhang
  et~al\mbox{.}}{2020}]%
        {zhang2020efficient}
\bibfield{author}{\bibinfo{person}{Wangda Zhang}, \bibinfo{person}{Mengdi Lin},
  {and} \bibinfo{person}{Kenneth~A Ross}.} \bibinfo{year}{2020}\natexlab{}.
\newblock \showarticletitle{Efficient Search over Genomic Short Read Data}. In
  \bibinfo{booktitle}{\emph{SSDBM}}.
\newblock


\end{thebibliography}
\end{document}